\documentclass[final,5p,times,twocolumn]{elsarticle}
\usepackage{amsfonts,amsmath,amssymb,bm}
\usepackage{graphicx} 
\usepackage{color}

\biboptions{sort&compress}

\def\noeq#1{(\ref{#1})}
\def\1eq#1{Eq.~(\ref{#1})}

\def\2eqs#1#2{Eqs.~(\ref{#1}) and~(\ref{#2})}
\def\3eqs#1#2#3{Eqs.~(\ref{#1}),~(\ref{#2}) and~(\ref{#3})}

\def\fig#1{Fig.~\ref{#1}}

\def\s#1{{\scriptscriptstyle #1}}

\def\rgl{\rho_\s{\hspace{-.03cm}\Delta}}
\def\rgh{\rho_\s{\hspace{-.03cm}D}}
\def\rD{\rho_\s{\mathcal D}}
\def\rghdr{\rho_\s{\hspace{-.03cm}F}}
\def\DN{{\cal D}}
\def\Drec{{\mathcal D}_{\hspace{-0.05cm}\rho}}
\def\mgl{m_\s{\mathrm{gl}}}
\def\hmgl{\widehat{m}_\s{\mathrm{gl}}}
\def\Zgl{Z_\s{\mathrm{gl}}}
\def\Zgh{Z_\s{\mathrm{gh}}}

\def\qp#1{{\mathfrak q}_{#1}}

\begin{document}

\begin{frontmatter}

\title{Spectral functions of confined particles}

\author[ECT]{Daniele Binosi}
\address[ECT]{European Centre for Theoretical Studies in Nuclear
Physics and Related Areas (ECT*) and Fondazione Bruno Kessler, \\Villa Tambosi, Strada delle
Tabarelle 286, 
I-38123 Villazzano (TN)  Italy}

\author[GU]{Ralf-Arno Tripolt}
\address[GU]{Institute for Theoretical Physics, Goethe University, Max-von-Laue-Str.~1, D-60438 Frankfurt am Main, Germany}

\date{17 April 2019}

\begin{abstract}

We determine the gluon and ghost spectral functions along with the analytic structure of the associated propagators from numerical data describing gauge correlators at space-like momenta obtained by either solving the Dyson-Schwinger equations or through  lattice simulations. Our novel reconstruction technique shows the expected branch cut for the gluon and the ghost propagator, which, in the gluon case, is supplemented with a pair of complex conjugate poles. Possible implications of the existence of these poles are briefly addressed.

\end{abstract}

\end{frontmatter}


\section{Introduction} 

As a consequence of the self-interactions of gauge fields, SU($N_\s{\!C}$) non-Abelian gauge theories are asymptotically free, {\it i.e.}, the coupling constant which controls the strength of their interactions decreases as the momentum scale increases~\cite{Gross:1973id,Politzer:1973fx}. This characteristic behavior persists even when matter particles are added, provided that their number $N_\s{\!f}$ is less than a critical value ($11 N_\s{\!C}/2$ at leading order, which is further reduced by non-perturbative effects, see, {\it e.g.}~\cite{Binosi:2016xxu}). This is indeed the case for Quantum Chromo-Dynamics (QCD), the theory thought to describe the strongly interacting sector of the Standard Model, in which one has $N_\s{\!C}=3$ (resulting in 8 adjoint gluons) and $N_\s{\!f}=6$ (corresponding to 6 fundamental quark flavors). The converse is also true: the QCD interaction strength grows with the separation between gluons and quarks. The theory is confined: colored states, like isolated gluons and/or quarks, do not appear in its spectrum.

Lacking direct access through experiments, confined particles have to be studied by theoretical {\it ab-initio} means. The quantity of choice here is their 2-point correlation function (the so-called propagator), which, in the last decade, has been extensively studied through both discrete (lattice regularized)~\cite{Cucchieri:2007md,Sternbeck:2007ug,Bogolubsky:2009dc,Ayala:2012pb,Duarte:2016iko} as well as continuum~\cite{Cornwall:1981zr,Aguilar:2008xm,Boucaud:2008ky,Dudal:2008sp,Fischer:2008uz,Tissier:2010ts,Maas:2011se,Strauss:2012dg,Huber:2012kd,Aguilar:2013xqa,Quandt:2013wna,Blum:2014gna,Cyrol:2016tym,Huber:2017txg,Huber:2018ned,Aguilar:2018csq} methods. When the theory is quantized in the Landau ({\it viz.}~a covariant~\cite{Bicudo:2015rma,Aguilar:2015nqa,Huber:2015ria,Capri:2015nzw,Siringo:2018uho}) gauge, and only the space-like Euclidean momentum region $q^2\geq0$ is considered, both methods agree: in the case of the (transverse) gluon propagator $\Delta_{\mu\nu}(q)=(g_{\mu\nu}-q_\mu q_\nu/q^2)\Delta(q)$ they describe a scalar cofactor $\Delta(q)$ that saturates in the deep infrared (IR) to a constant non-vanishing value, $\Delta(0)=1/\mgl^2>0$, which can be interpreted as being due to the dynamical generation of an effective gluon mass~\cite{Cornwall:1981zr,Binosi:2009qm,Aguilar:2011ux,Binosi:2012sj}. For the ghost particle (which appears as a consequence of the gauge fixing procedure) it is rather the dressing function $F(q)=q^2 D(q)$, with $D$ the ghost propagator, that saturates: the ghost remains non perturbatively massless~\cite{Boucaud:2008ky}.

\begin{figure*}[!t]
	\includegraphics[scale=0.37]{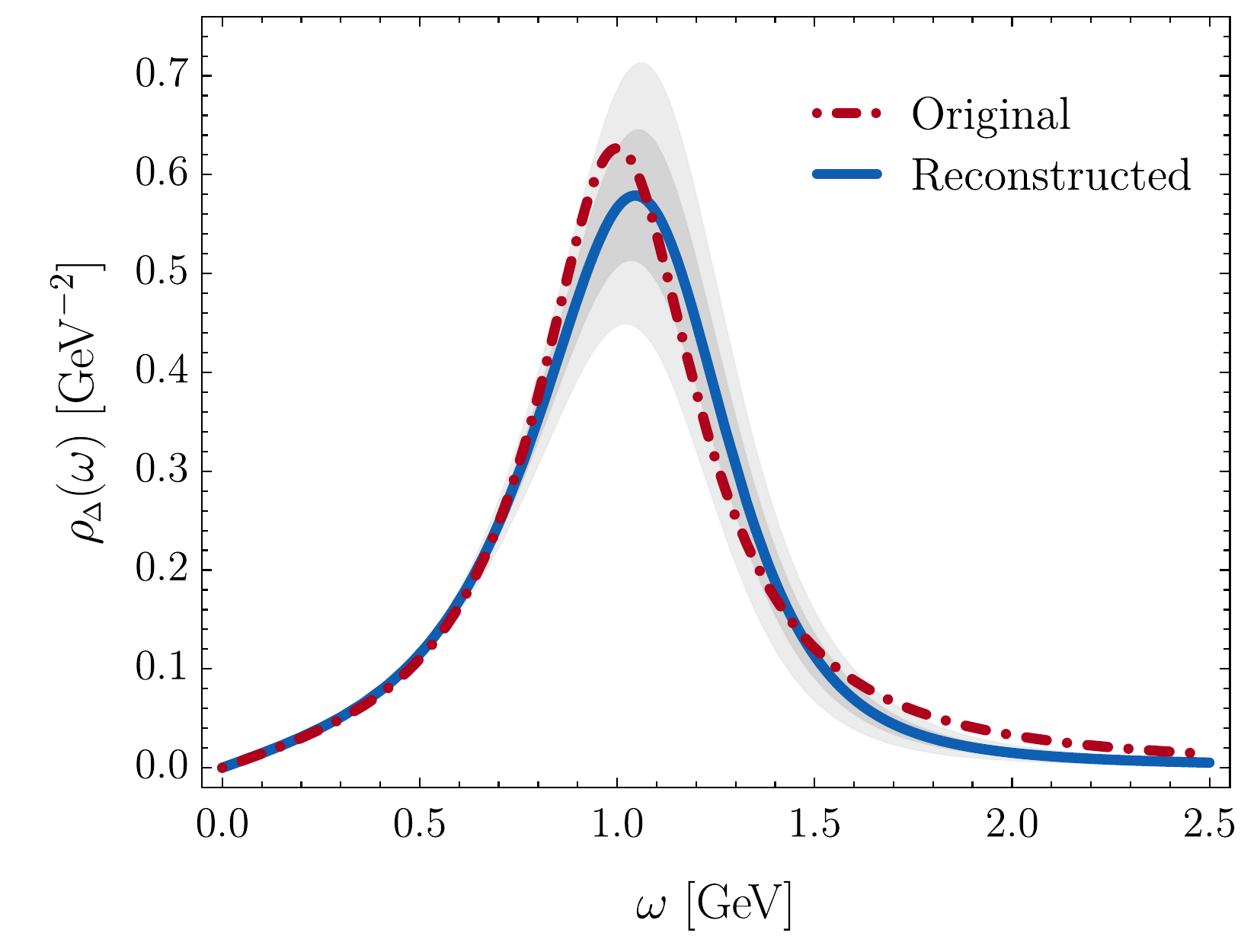}
	\includegraphics[scale=0.37]{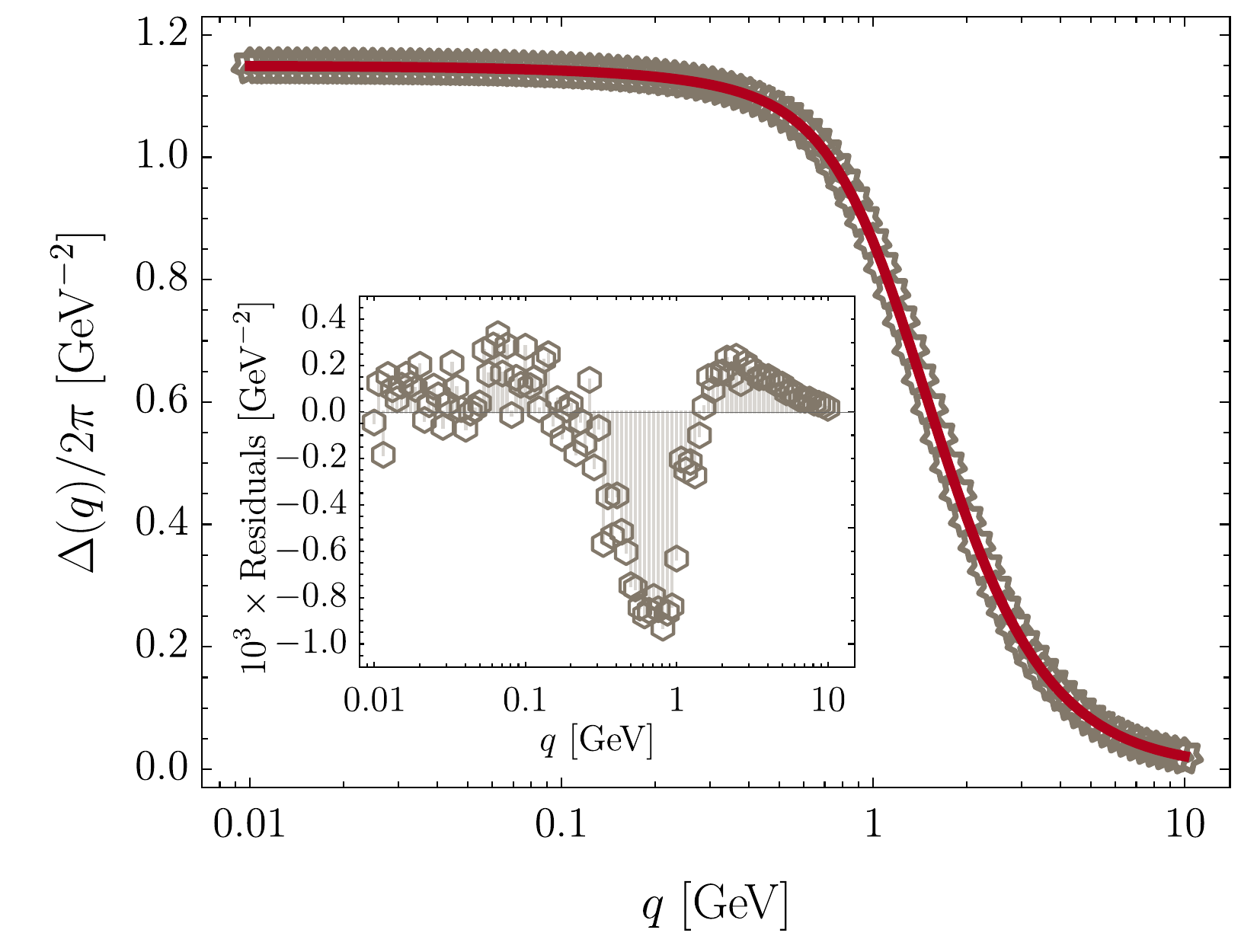}
	\includegraphics[scale=0.37]{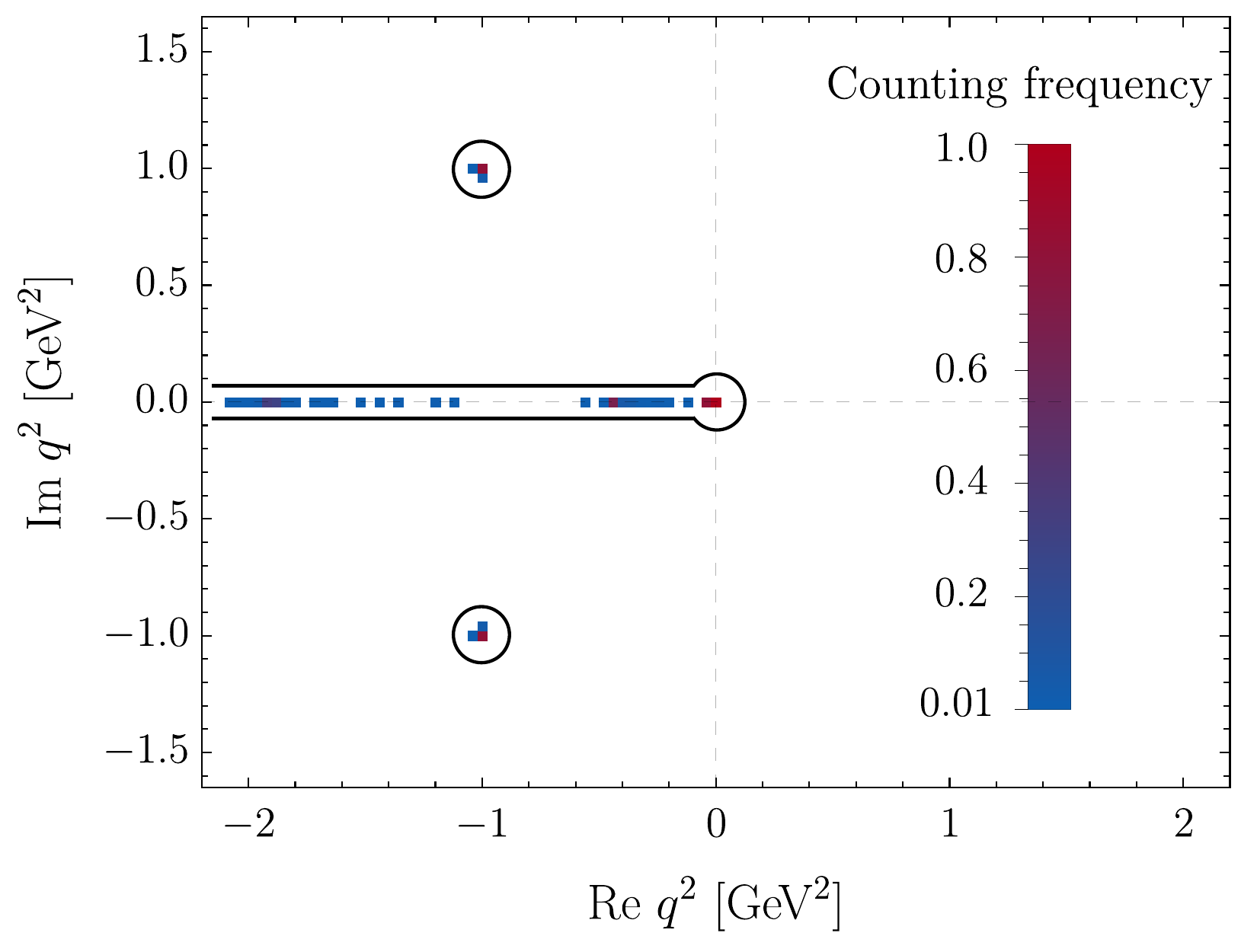}
	\caption{\label{fig:BW_prop} Left: Reconstructed spectral function or the Breit-Wigner with noise test case; light (dark) gray areas correspond to one (two) $\sigma$ confidence regions (see Appendix).  Middle: the propagator curve plotted obtained from the smallest norm spectral function. Right: histogram of all poles of the SPM propagators that lie in the left half-plane, thus omitting Froissart doublets in the right half-plane; the counting frequency is defined as $N/N_{\text{max}}$, where $N$ is the number of poles in a given bin and $N_{\text{max}}$ the maximum number of poles per bin found in the histogram; the branch cut at $q^2\leq 0$ and the poles at $\qp{1}^{2}=-1\pm i$ are correctly identified. The poles used in~\noeq{Lehmann} are only those with residues above a certain threshold, see text for details.}
\end{figure*}

Knowledge on the full propagator is often equalled with having solved the underlying theory. It is the central object in the calculation of equilibrium as well as dynamical observables with applications ranging from the determination of the hadron and glueball spectrum to the calculation of transport coefficients of the quark gluon plasma. The determination of the full propagator and its analytic structure in the complex plane is therefore indispensable. 

But how does the analytic structure of these correlators look like? Consider for simplicity the pure glue theory in which the only dynamical scale is $\Lambda_{\s{\mathrm{QCD}}}$. Then, perturbation theory implies that, whenever the condition $\widehat q^2\equiv q^2/\Lambda^2_{\s{\mathrm{QCD}}}\gg1$ is met, one has at leading order~\cite{Fischer:2003zc} $\Delta(q)\sim \Zgl/q^2(\log\widehat q^{\,2})^\gamma$ and $D(q)\sim \Zgh/q^2(\log\widehat q^{\,2})^\delta$ where $\Zgl$ and $\Zgh$ are dimensionless renormalization constants and $\gamma$ and $\delta$ are the (one-loop) anomalous dimensions (with $\gamma=13/22$ and $\delta=9/44$ for $N_\s{\!f}=0$, which 
	will be the case from now on). 
Accounting for the perturbative logarithmic running in the ultraviolet (UV) momentum region, requires the presence of a branch cut structure for real and time-like $q^2$ momenta. The presence of this cut, without any additional singularities in the complex plane, would in turn allow for the usual K\"all\'en-Lehmann representation by which the full propagator is expressed as an integral over the (free) propagator with a spectral density depending on the frequency~$\omega$. 

In fact, owing to the axioms of {\it local} quantum field theory \cite{Haag1996}, any 2-point correlation function must be an analytic function in the cut complex $q^2$-plane with singularities along the time-like real axis only~\cite{Alkofer:2000wg}; for any other singularity structure to be present, {\it e.g.}, (simple) poles, one or more of these axioms, typically local space-like commutativity, {\it i.e.},~causality, needs to be relaxed
As has been discussed in~\cite{Habel:1989aq,Habel:1990tw,Stingl:1994nk}, however, although causality might be violated at the level of the 2-point functions in such cases, the corresponding $S$-matrix remains both causal and unitary. One might also argue~\cite{Maas:2011se} that strict locality may be too strong a requirement for physical theories displaying confinement, in view of non-local complications like, {\it e.g.}, the Gribov-Singer ambiguity~\cite{Gribov:1977wm,Singer:1978dk}. Furthermore, several phenomenological models predict the appearance of complex simple poles (which are bound to materialize in conjugate pairs, for the propagator is real at space-like momenta) in the gluon propagator~\cite{Dudal:2008sp,Sorella:2010it,Siringo:2015wtx}, see also \cite{Gribov:1977wm,Stingl:1985hx,Zwanziger:1989mf}. Further evidence for their existence has been found in the quark sector where the Nakanishi representation~\cite{Nakanishi:1963zz,Nakanishi:1969ph,Nakanishi1971} perfectly describes 
solutions of the quark gap equation~\cite{Chang:2013pq} using the most sophisticated quark-gluon kernel available~\cite{Qin:2011dd,Binosi:2014aea} (see also~\cite{Williams:2015cvx,Cyrol:2017ewj}). 

Thus, to ascertain whether or not such complex conjugate poles appear in the analytic structure of QCD propagators, a method that assumes no (or at least as minimal as possible) prior knowledge on such structures is needed. There are several numerical continuation methods available in the literature that aim at obtaining the best possible reconstruction of spectral functions~\cite{Jaynes1957,Jaynes1957a,BackusGilbert1968,G.Backus1970,HobsonLasenby1998,PressTeukolskyVetterlingEtAl2002,Alkofer:2003jj,Burnier:2013nla,Haas:2013hpa,Dudal:2013yva,Christiansen:2014ypa,Dudal:2019gvn}; however, as all these reconstruction techniques rely on the standard K\"allen-Lehmann-type spectral representation, they ignore the possible presence of additional structures in the complex plane. The purpose of this letter is to describe a novel numerical analytic continuation tool allowing for this possibility, and apply it to the available continuum and lattice data for the gluon and ghost propagators in order to obtain their respective spectral functions together with the corresponding analytic structure. 

\section{Generalized spectral representation and reconstruction method} 

In the presence of $n$ complex conjugate pairs of simple poles located at $\qp{i}$, $\qp{i}^{*}$ ($i=1,..,n$) and with residue $R_i$, $R_i^*$, the usual K\"all\'en-Lehman spectral representation gets generalized to
	\begin{align}
		{\cal D}(q)&=\frac1\pi\int_0^\infty\!{\mathrm d}\omega\,\frac{\omega\rD(\omega)}{\omega^2+q^2}+\sum_i \frac{R_i}{q^2-\qp{i}^2}+\sum_i \frac{R^*_i}{q^2-\qp{i}^{*2}},
		\label{Lehmann}
	\end{align}
with
	\begin{align}
		\rD(\omega)&=2\ \mathrm{Im}\ {\cal D}(-i(\omega+i\epsilon)),
		\label{SF}
	\end{align}
where $q$ refers now to the energy-component and $\rD$ is the (non-positive definite) spectral density defined at real frequencies $\omega$, with $\epsilon\to0^+$~
\cite{Siringo2017, Hayashi:2018giz}. 
As $D(q)\sim 1/q^2$ in the IR, for the ghost particle we will study the spectral function associated to its dressing function $\rghdr$ rather than $\rgh$, the two being related by $\rgh=F(0)\delta'(\omega)-\rghdr/\omega^2$~\cite{Dudal:2019gvn}.
We also note that assuming $q^2\Delta(q)\to0$ for $q^2\to\infty$, as required by perturbation theory, implies a modified version of the usual Oehme-Zimmermann superconvergence (OZS) relation~\cite{Oehme:1979bj,Oehme:1990kd}, which now becomes $\int_0^\infty\!{\mathrm d}\omega\,\omega\rD(\omega)+4\pi\sum_i\mathrm{Re}\,R_i=0$.
\begin{figure*}
	\includegraphics[scale=0.37]{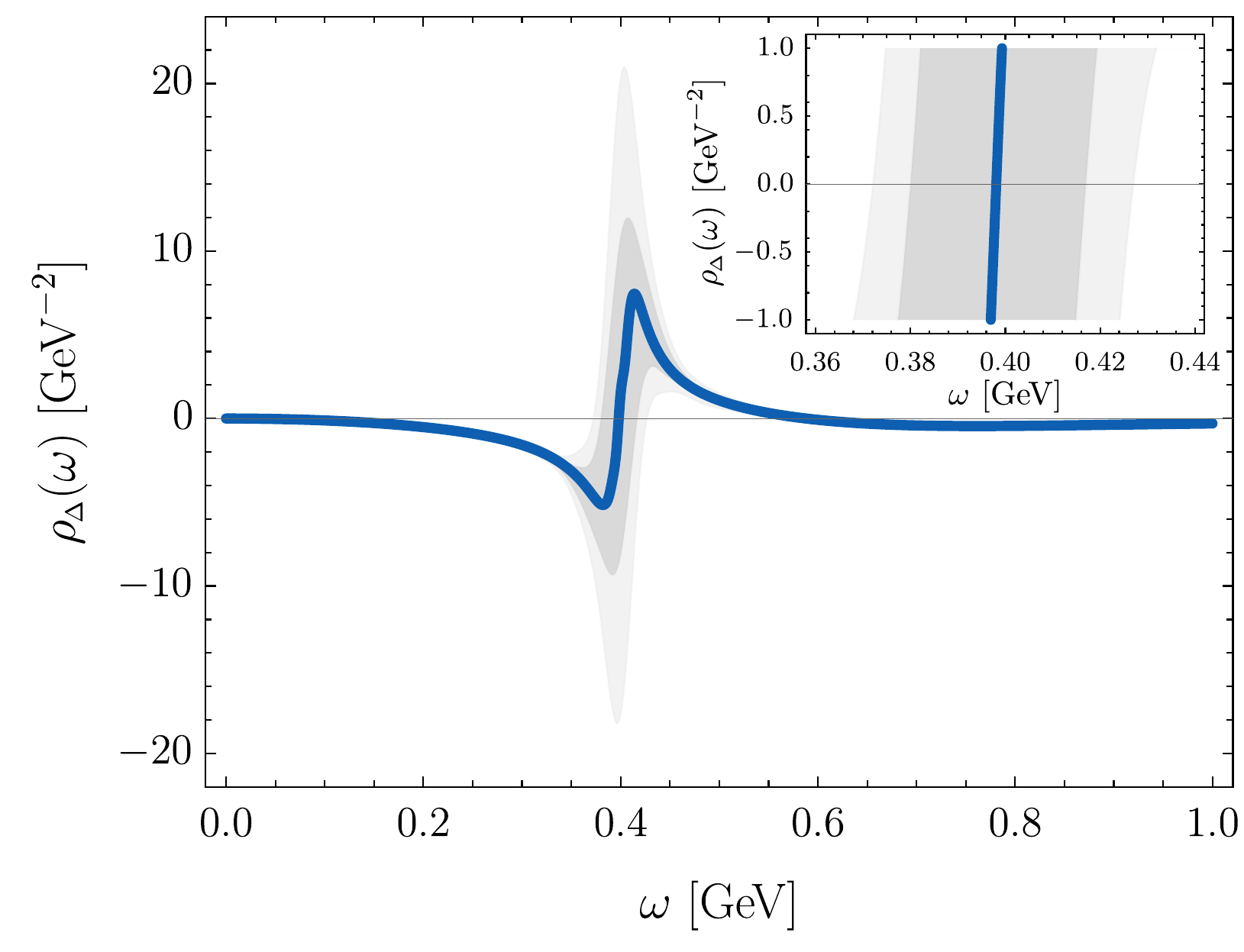}
	\includegraphics[scale=0.37]{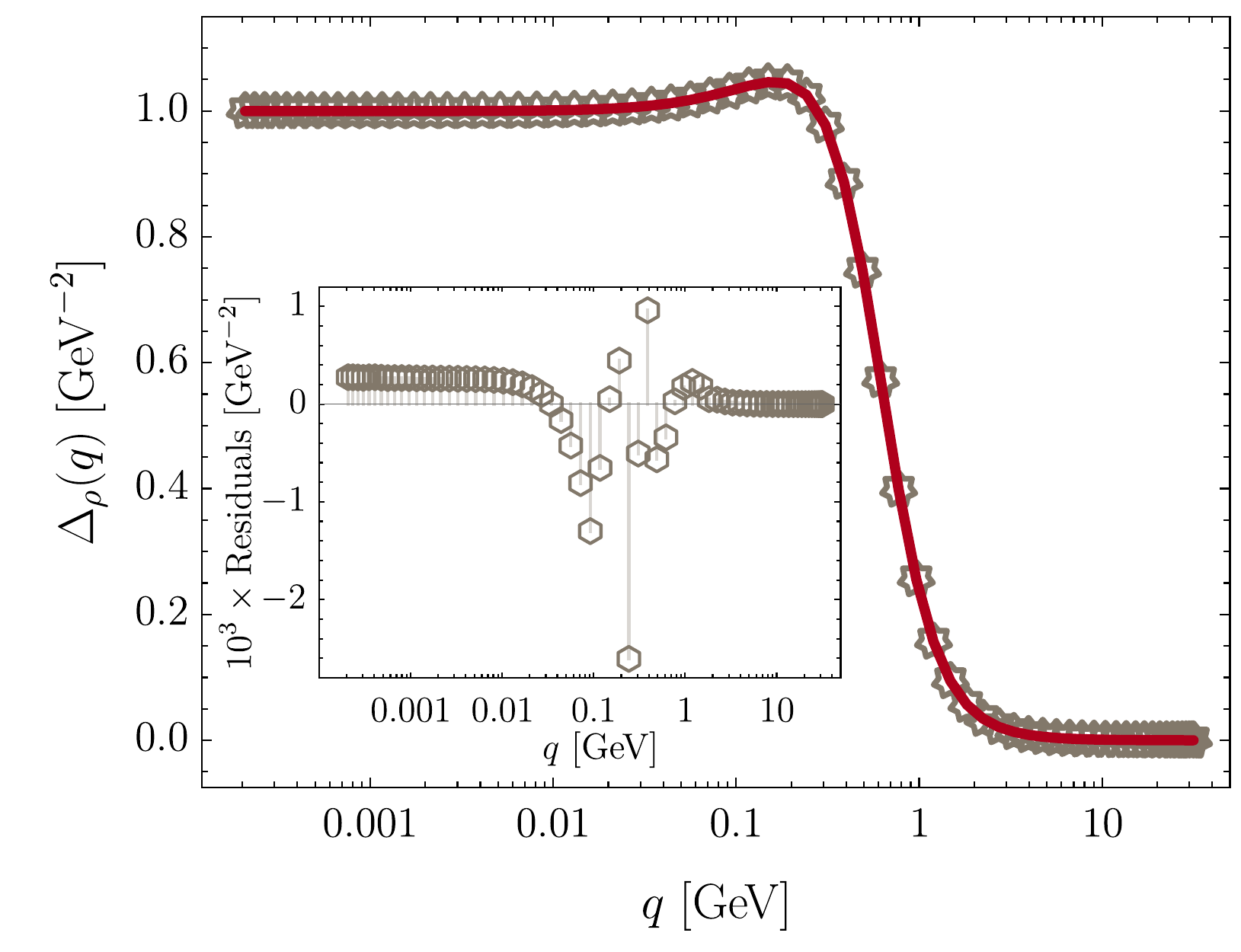}
	\includegraphics[scale=0.37]{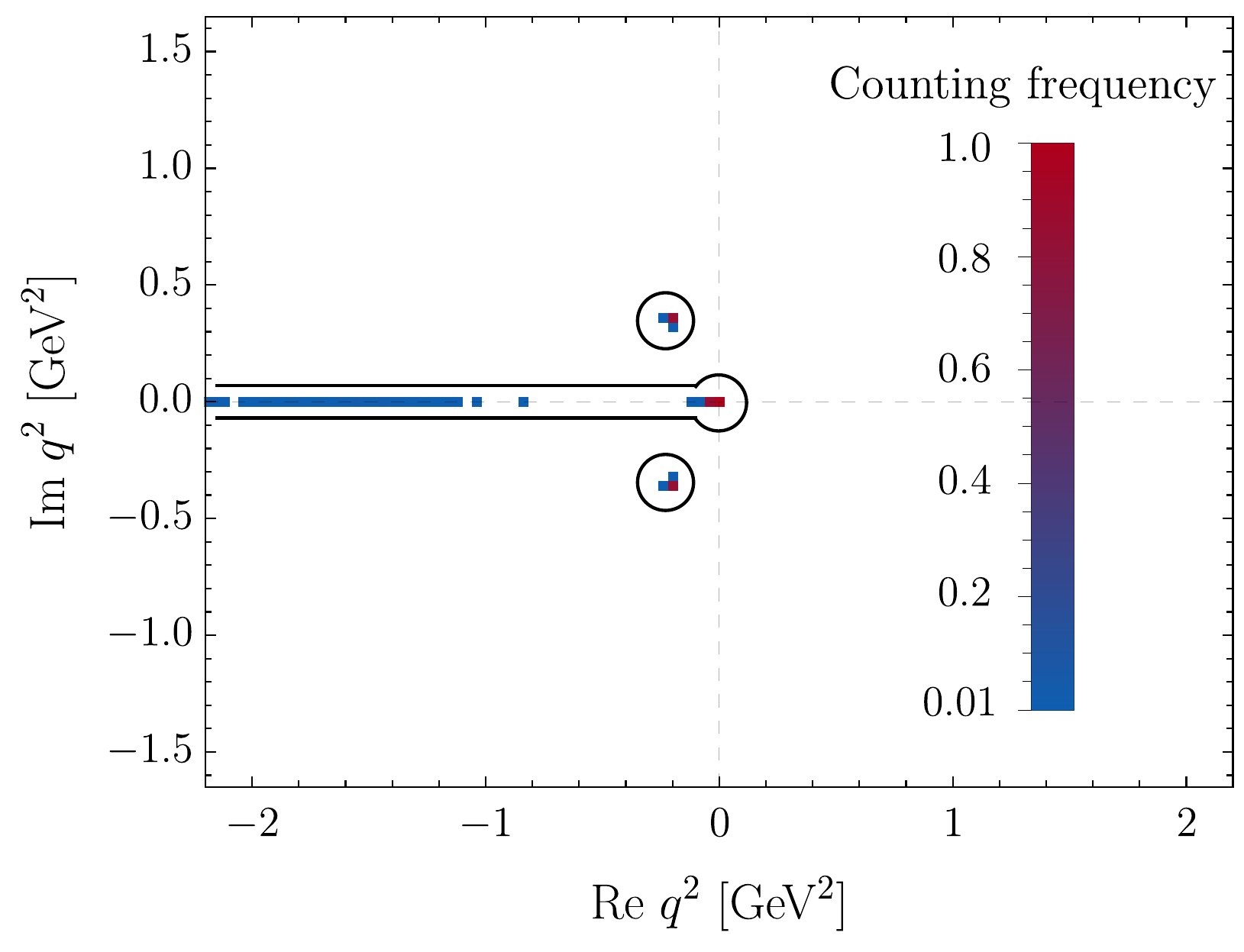}
	\includegraphics[scale=0.37]{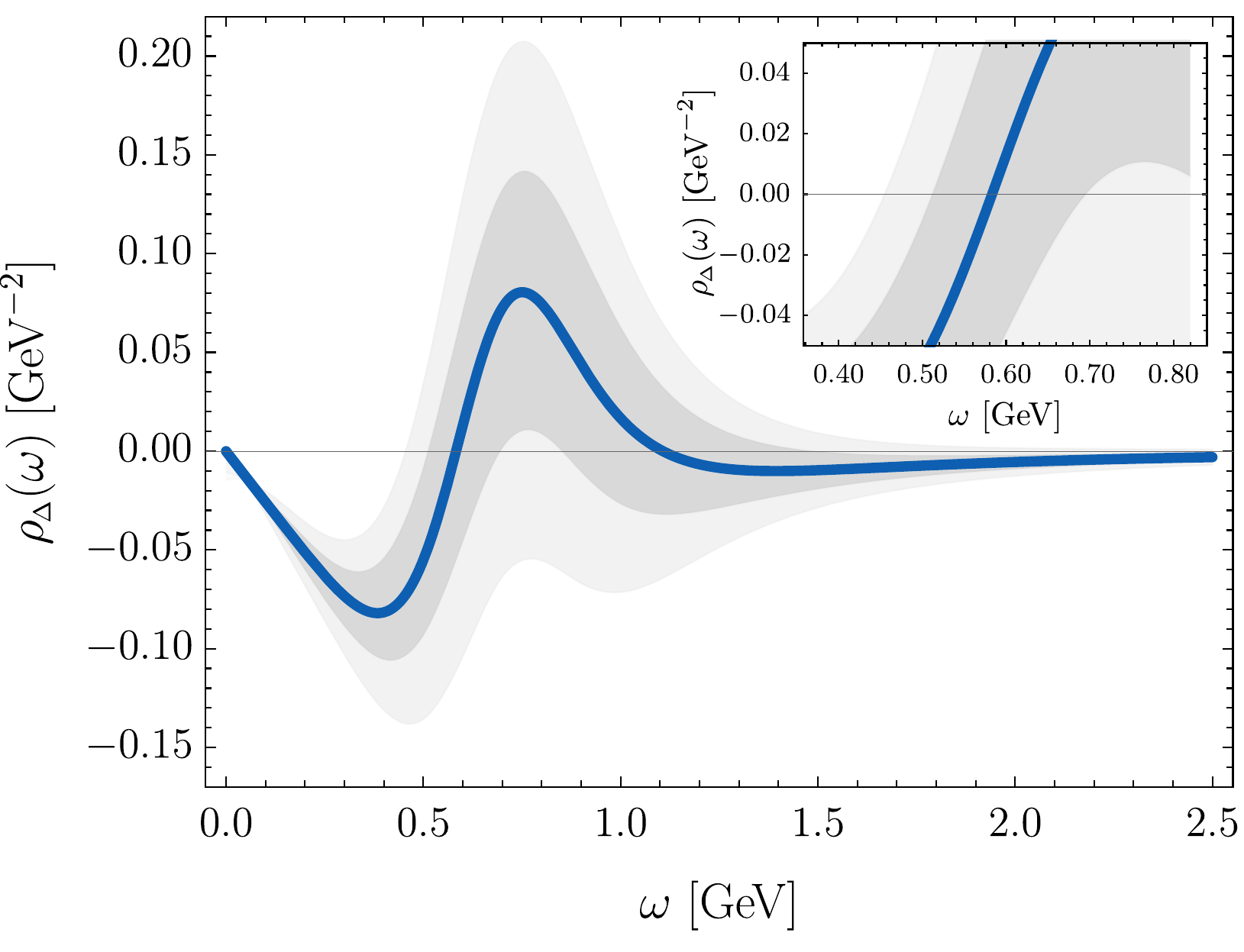}
	\includegraphics[scale=0.37]{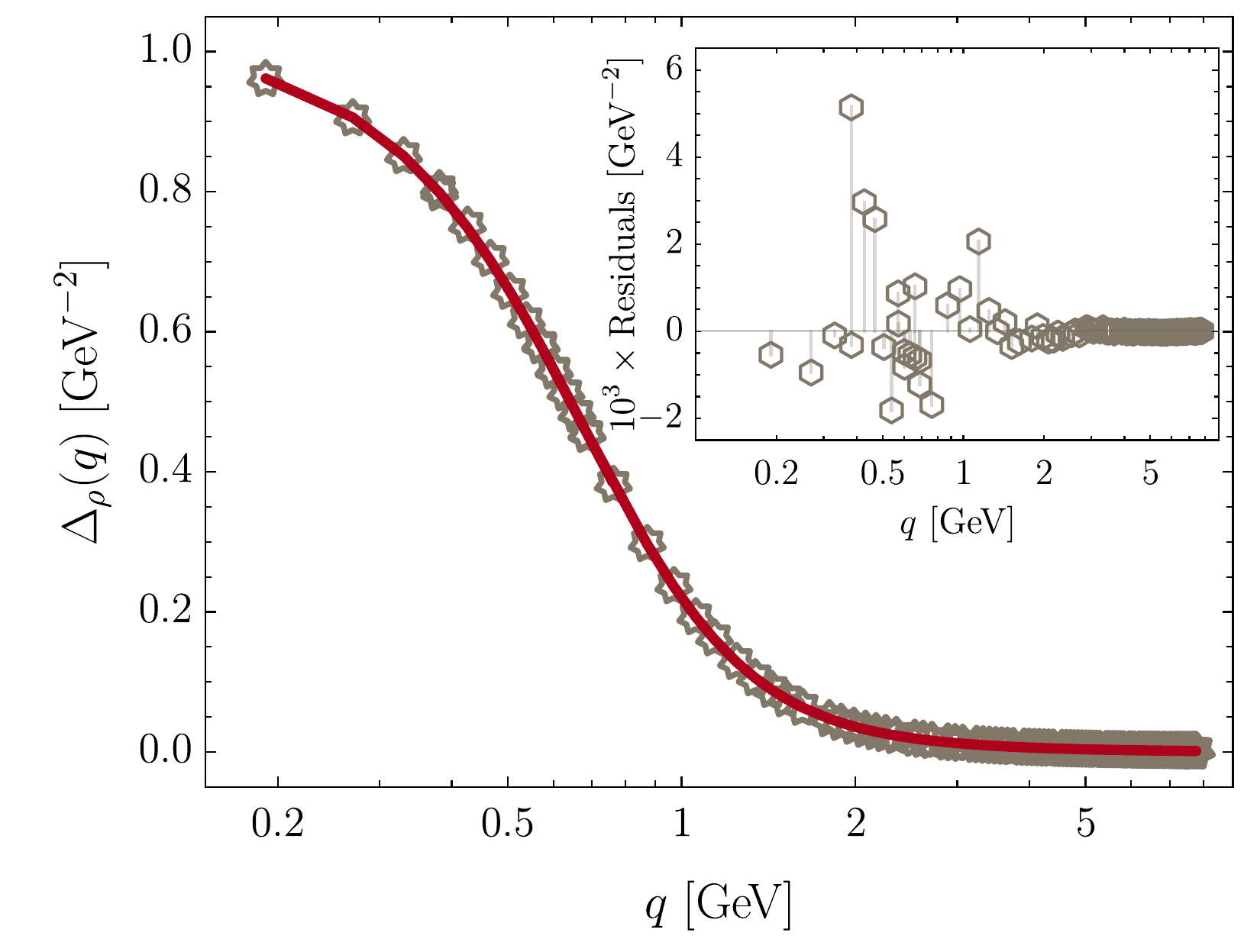}
	\includegraphics[scale=0.37]{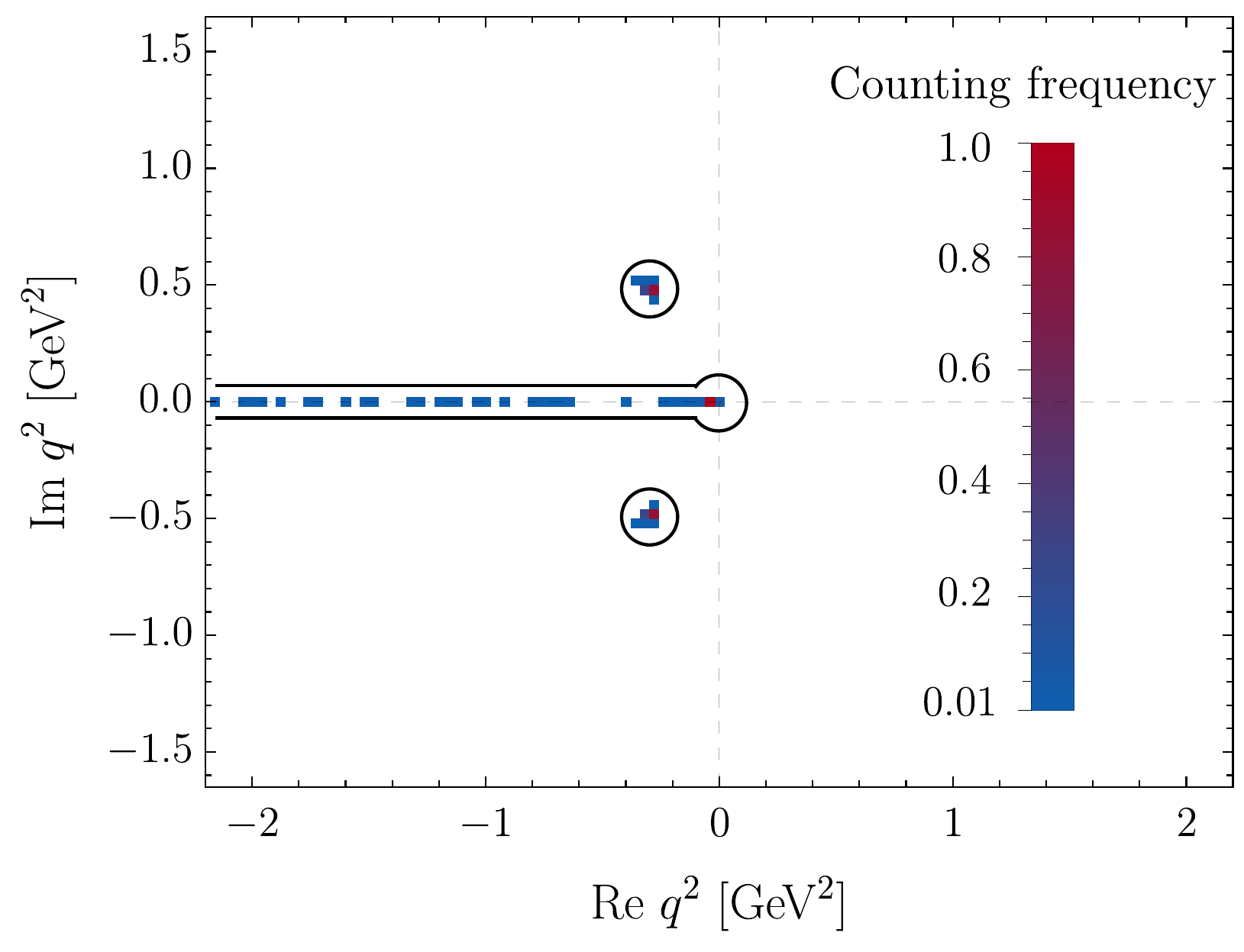}
	\caption{\label{fig:gl-CF-OO}
	As before, but for the gluon DS (top) and lattice (bottom) data. The presence of a pair of complex conjugate poles is clearly visible in the right panels for both cases.}
\end{figure*}

The proposed reconstruction algorithm is a variant of the Schlessinger Point Method (SPM) which is based on a rational-fraction representation similar to Pad\'{e} approximants \cite{Schlessinger1966}. Given a set of $N$ input points $(q_i, y_i={\cal D}(q_i))$ one first constructs a continued fraction representation of the propagator:
\begin{equation}
	\DN(q) =\frac{y_1}{1+}\frac{a_1(q-q_1)}{1+}\frac{a_2(q-q_2)}{1+}\cdots\frac{a_\s{N-1}(q-q_\s{N-1})}{1},
	\label{eq:pade2}
\end{equation}
where the coefficients $a_i$ are recursively evaluated such that the function $\DN(x)$ acts as an interpolation through all the points, {\it i.e.}, $\DN(q_i)=y_i$, $i= 1,2,\dots, N$ (see~\cite{Schlessinger1966, Tripolt:2017pzb, Tripolt:2018xeo} for details). Once the function $\DN(q)$ is determined, its analytic continuation is defined as the meromorphic function obtained by allowing the originally real $q$ to take on complex values; the corresponding spectral function, $\rD$, can be then evaluated using~\1eq{SF}. The quality of the reconstructed spectral function can be assessed by inserting~$\rD$ in~\1eq{Lehmann} and comparing the resulting propagator $\Drec$ with the original one by means of a suitable norm, the simplest one being 
\begin{align}
    ||\Drec,{\mathcal D}||=\sum_{i=1}^N\frac{[\Drec(q_i)-y_i]^2}{y_i}.
    \label{eq:norm}
\end{align}
Notice that the OZS relation is not used to further constrain the spectral function since we focus on intermediate energies and the reconstructed spectral function by construction behaves asymptotically as $1/\omega$, giving rise to a divergent integral. 

In the following we will be dealing with data sets comprising $M>N$ values for the propagators ${\mathcal D}$ in the space-like Euclidean momentum region, from which we want to reconstruct the associated spectral functions. We are therefore confronted with the problem of which $N$ points should be chosen as our input. In fact, if the data were exact, {\it i.e.}, without any statistical error, any subset chosen would give, in general, the same result. In realistic situations, however, data do have errors, and some subsets of input points will give results that are closer to the correct one ({\it i.e.}, the one that would be obtained for exact data) than others. In order to identify such `optimal' subsets we have devised an algorithm comprising the following steps:
\begin{itemize}
    \item[({\it i})] Select a partition of $N$ random points chosen from the complete set of $M$ points $(q_i, y_i)$; 
    \item[({\it ii})] Use~\1eq{eq:pade2} to get the corresponding $\DN(q_i)$;
    \item[({\it iii})] Use~\1eq{SF} to calculate the spectral function; if there are known constraints, {\it e.g.}, on its asymptotic behavior and/or its positiveness (or lack thereof), this information can be used here to either accept or reject the spectral function;
    \item[({\it iv})] Reconstruct the propagator $\Drec$ from~\1eq{Lehmann}, taking only complex conjugate poles with a residue above a certain threshold into account, and evaluate its norm~\noeq{eq:norm}; 
    \item[({\it v})] Repeat steps ({\it i}) through ({\it iv}) for $L$ times and identify the input point $(q_j, y_j)$ that was used most often among the subset of $K$ functions with the smallest norm; 
    \item[({\it vi})] Repeat steps ({\it i}) through ({\it v}) $N-1$ times, keeping fixed the identified points $(q_j, y_j)$ until all $N$ optimal points have been selected. 
\end{itemize}

\begin{figure*}
	\includegraphics[scale=0.37]{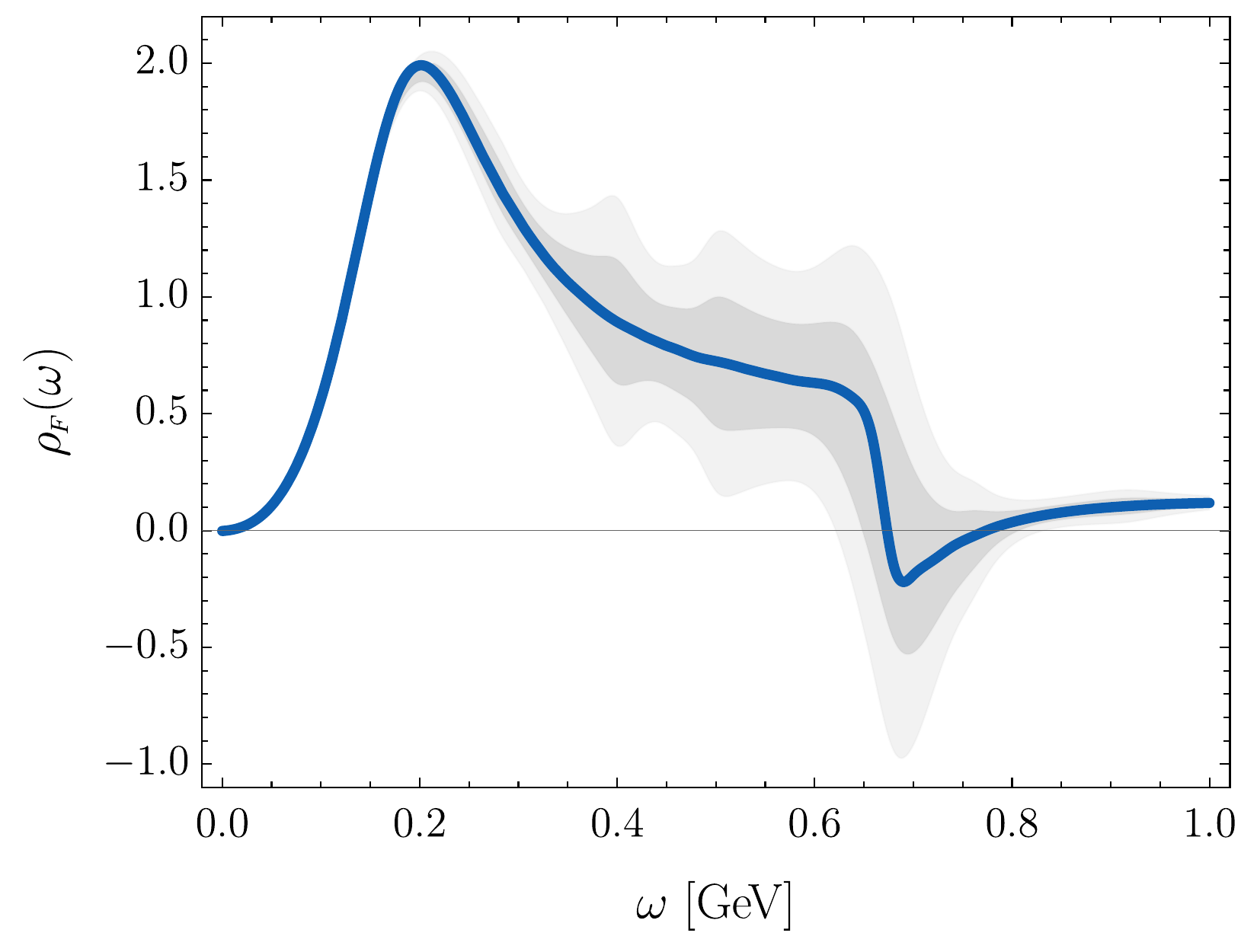}
	\includegraphics[scale=0.37]{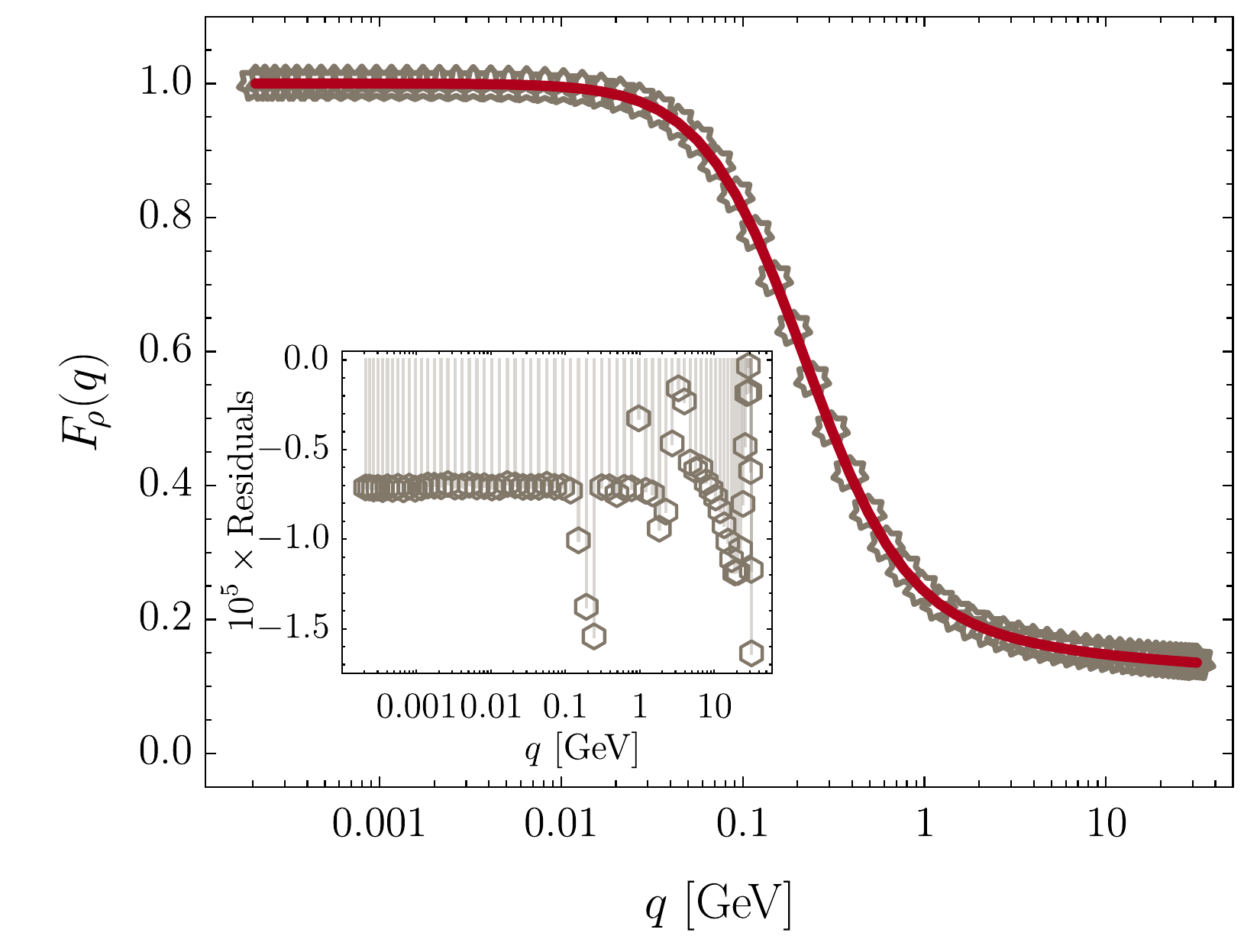}
	\includegraphics[scale=0.37]{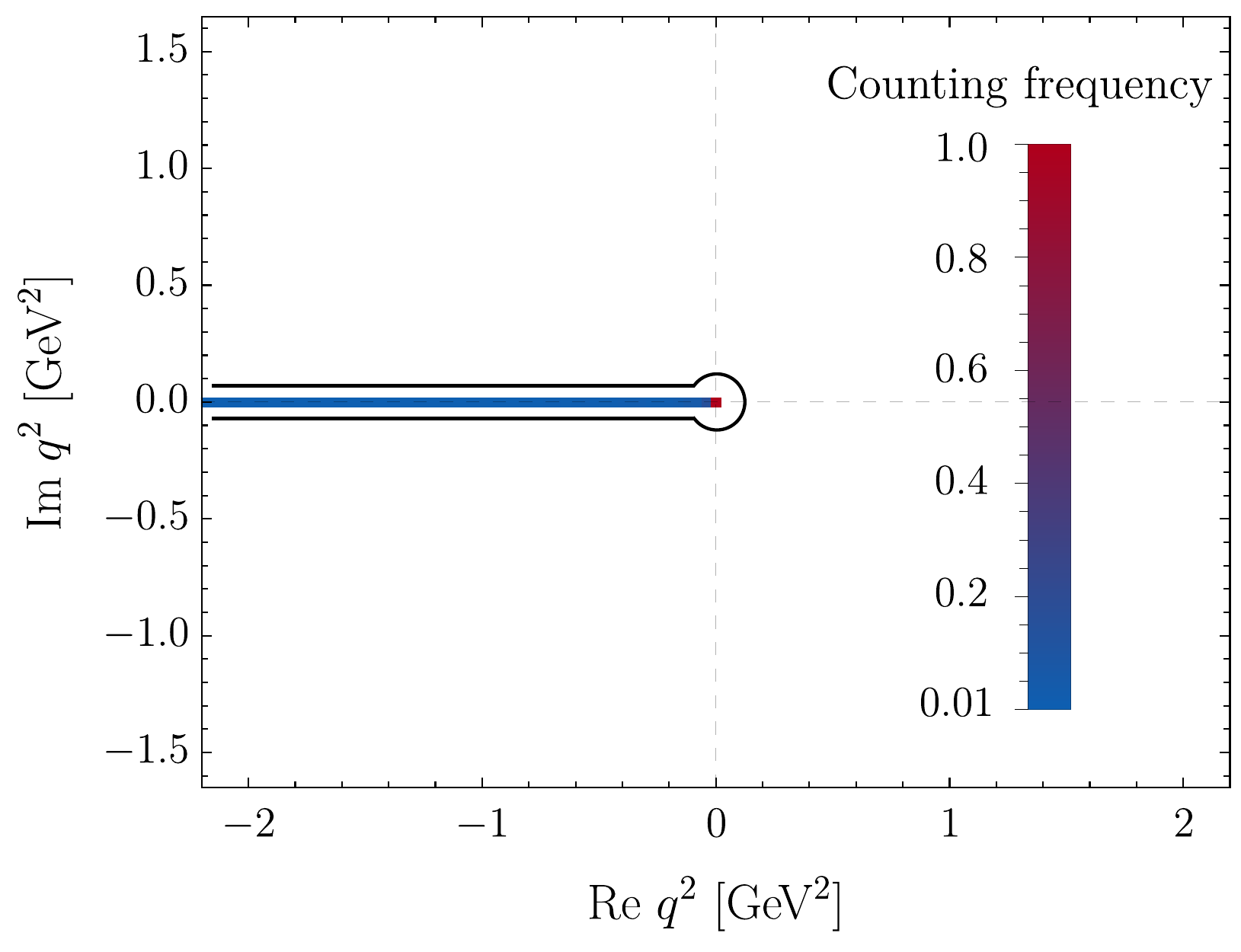}
	\includegraphics[scale=0.37]{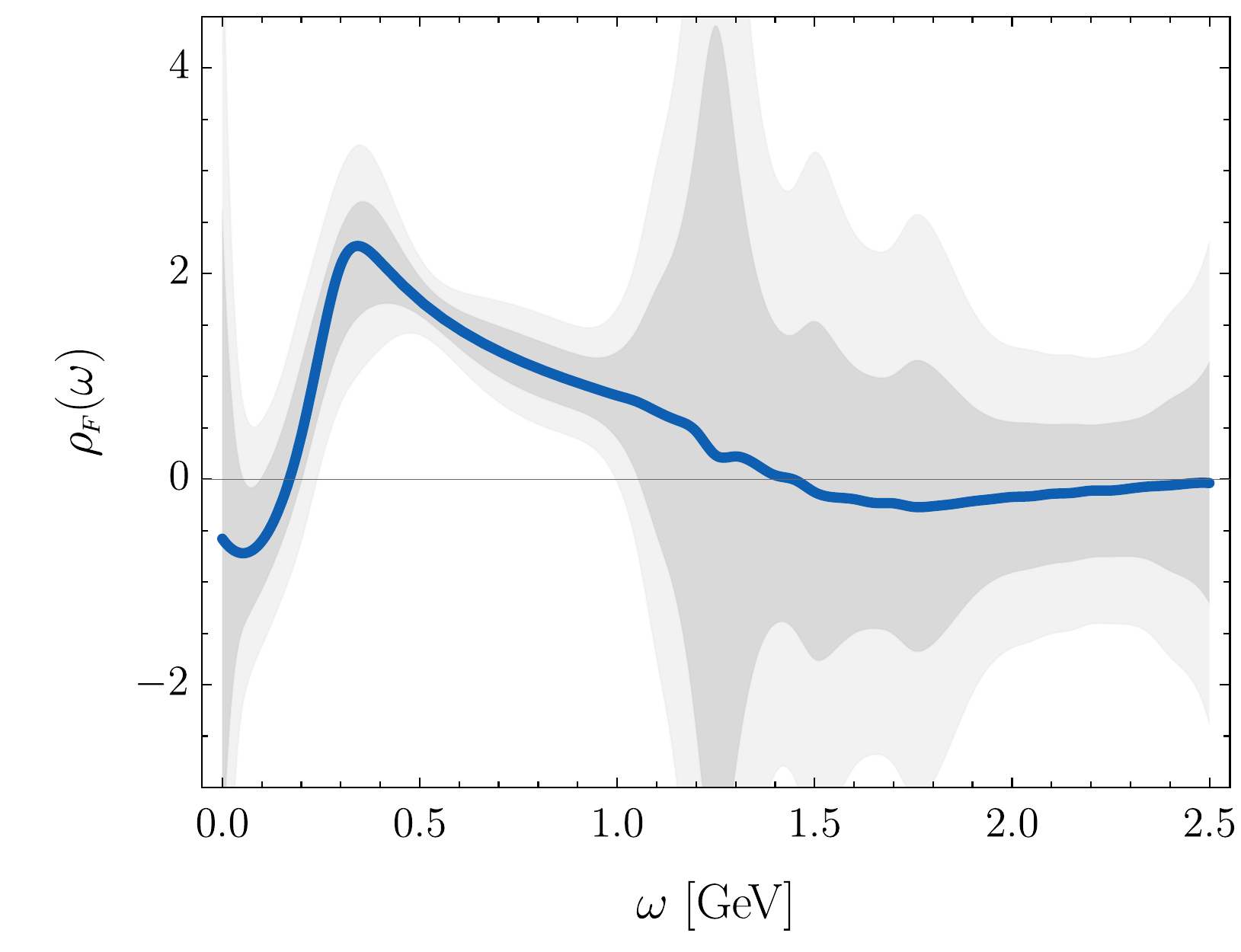}
	\includegraphics[scale=0.37]{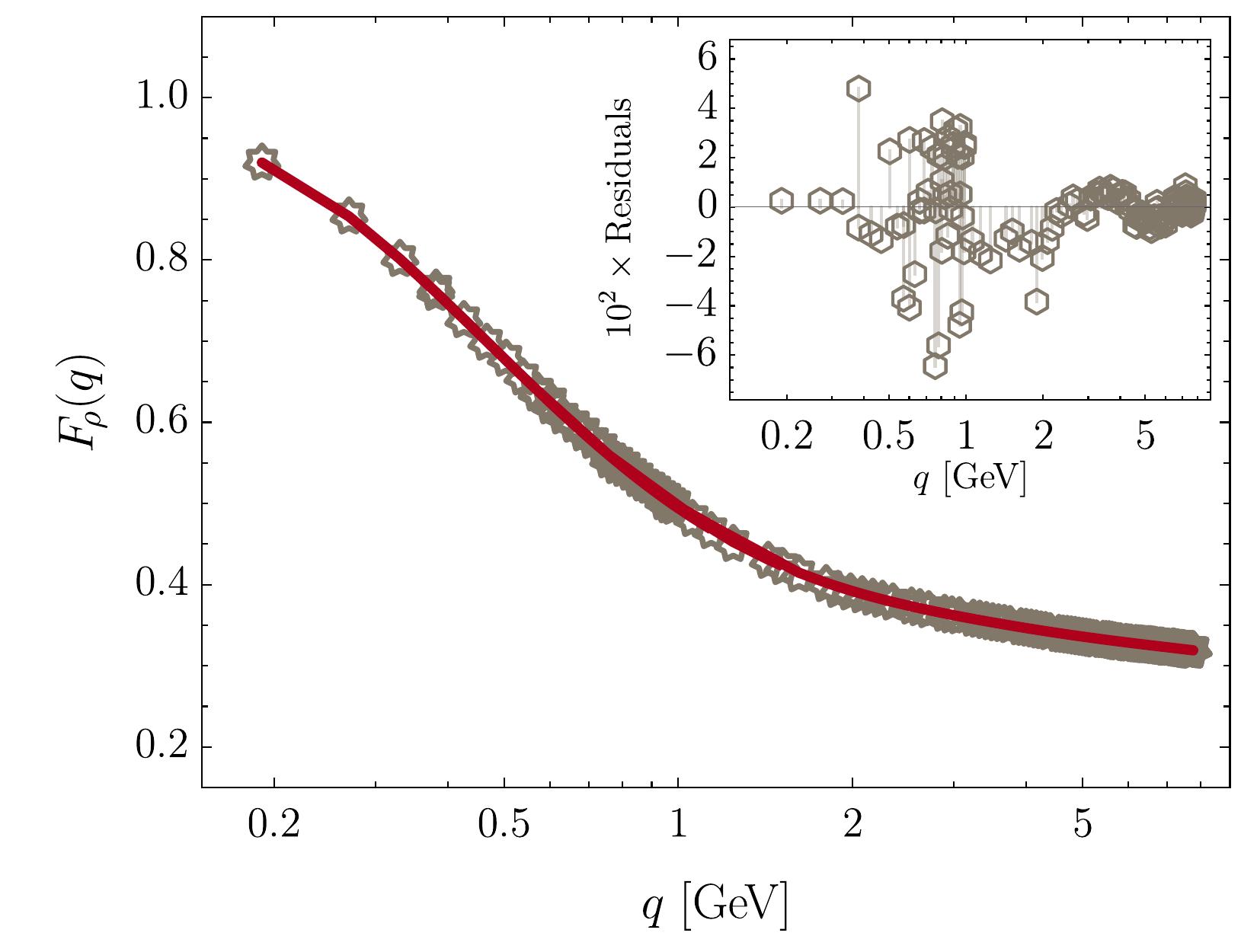}
	\includegraphics[scale=0.37]{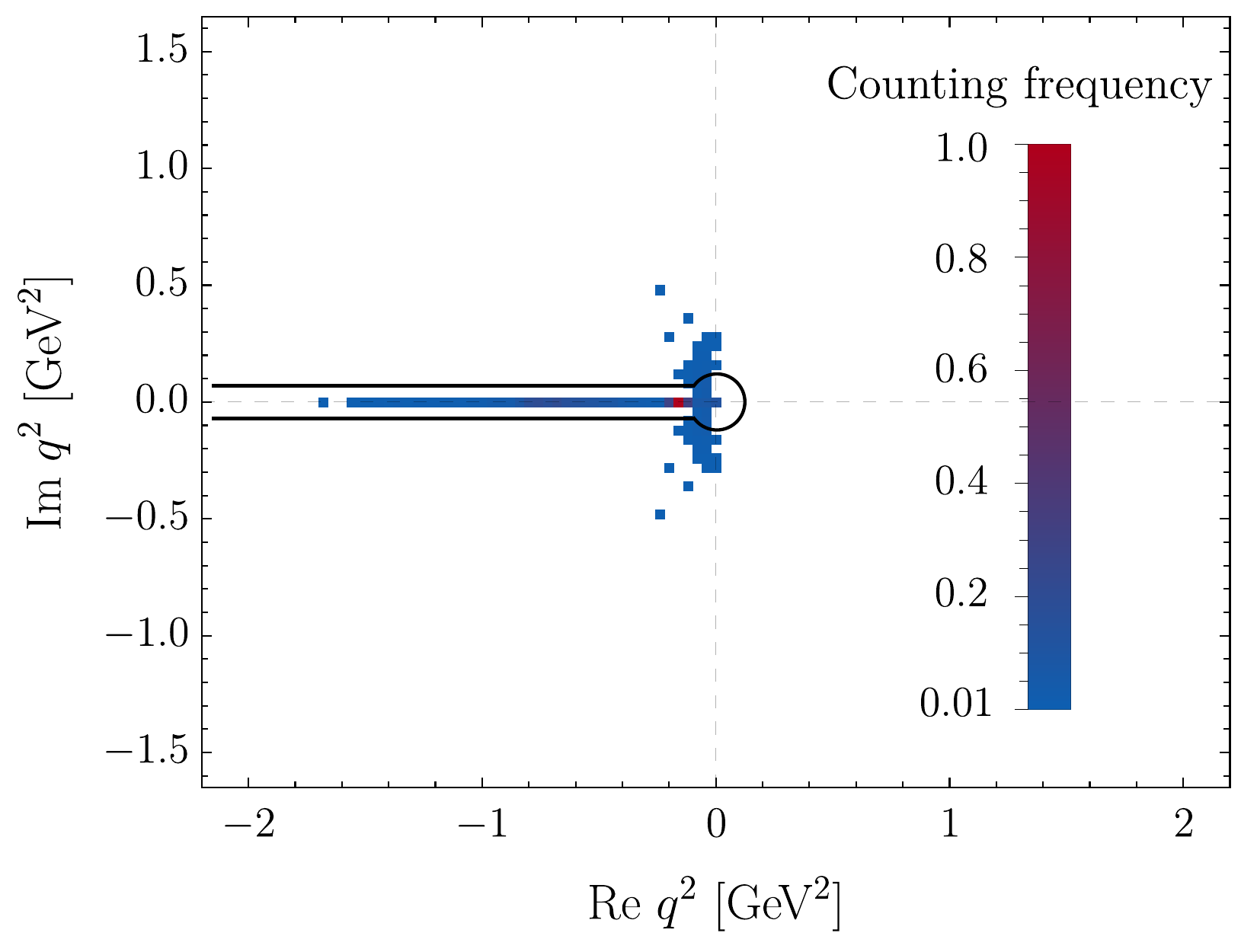}
	\caption{\label{fig:gh-CF-OO}
	As before, but for the ghost DS (top) and lattice (bottom) data. Notice the appearance of spurious poles in the lattice complex $q^2$-plane reconstruction due to the (lack of) precision of the used data.}
\end{figure*}

As described in $({\it iv})$, only poles with residues above a certain threshold are taken into account when evaluating \1eq{Lehmann}. This threshold is chosen such as to discard non-physical Froissart-doublet poles while at the same time keeping candidates for physically meaningful poles. In the present work, we use a threshold in the range between 0.1 and 2. Let us now briefly discuss the values chosen for the different parameters. To begin with we note that, if $\DN$ in~\1eq{Lehmann} is expressed as a rational fraction, $\DN = P/Q$, $P$ and $Q$ will be polynomials of order $(N-1)/2$ ($P$ and $Q$) for an odd number of input points, and of order $N/2-1$ ($P$) and $N/2$ ($Q$) for an even number of input points. Even though knowing that the propagator vanishes in the UV leads to the natural choice of $N$ even, one cannot hope in any case to numerically reproduce its asymptotic behavior correctly based on noisy input data; and, in fact, in the intermediate energy range considered here, we checked that our results are stable irrespectively of $N$ being even or odd. What matters is that there seems to be always an optimal number of input points, call it $N_*$:  for fewer input points we found that the reconstruction  depends strongly on the number of points, whereas a larger number gives rise to numerical instabilities due to a loss of accuracy when calculating the coefficients of the continued fraction~(2) (the C$++$ code developed uses the GMP library for arbitrary precision arithmetic on floating point numbers). For all datasets we have found that choosing around 50 input points always works, and therfore set $N_*=50$. For the remaining parameters we have set $L=4000$ and $K=200$ which offered a good compromise between execution speed and results precision. $M$ varies among the different data sets from a minimum of $69$ (DS gluon case) to a maximum of $148$ (lattice ghost case); finally we effectively set $\epsilon$ appearing in~\1eq{SF} equal to zero.

As step ({\it i}) of the algorithm is intrinsically random, every run of the algorithm will result, in general, in a different set of optimal points. Typically, we had 500-1000 full runs per case (performed on the Goethe-HLR cluster), out of which we selected only the ones with a norm below a suitable threshold, set at 1.5 times the minimum value reached. All solutions are then resampled in order to construct the mean and the standard deviation, thus furnishing respectively the best curve and the one and two $\sigma$ confidence levels (gray error bands in the figures).

We proceed with a model application of the algorithm in order to demonstrate its capabilities. For this purpose, we generate mock propagator data consisting of 100 points $(q_i,y_i)$ in the range $q_i\in[0.01,10]$ GeV, using~\1eq{Lehmann} with a Breit-Wigner spectral function 
$2\Gamma\omega/[\pi((\omega^2-\Gamma^2-M^2)^2+4\Gamma^2\omega^2)]$
with $M=4\Gamma=1$~GeV and supplemented with a pair of complex-conjugate poles located at $\qp{1}^2=-1\pm i$ GeV$^2$ with residue $R_1=1$. Finally, we add a statistical error to the input data through $y_i\to y_i(1+\varepsilon r_i)$ with $\varepsilon=10^{-3}$ and $r_i$ a random number drawn from a normal distribution with zero mean and unit standard deviation.
We note that only the central values of the data points are used for the reconstruction. The uncertainties are therefore taken into account effectively, as encoded in the scattering of the central values. The error range can in principle also be taken into account explicitly by generating data points within the given error bars.
As can be seen in~\fig{fig:BW_prop}, the method determines the correct $\rD$ and the corresponding propagator~$\Drec$, also accurately identifying the location of the poles in the complex $q^2$-plane, see also~Table~\ref{tab:polelocation}. The presence of a branch cut, to be expected on analytic grounds due to a square root term appearing when integrating the Breit-Wigner spectral function, is also visible as a sequence of poles on the negative real axis. Indeed, the reconstruction of the meromorphic structure of the function is possibly the most stable result that the algorithm gives: increasing the error rate $\varepsilon$ results in a loss of precision in locating peak position and height of $\rD$, but one will still get a reliable estimate on the cut and poles positions.  

\begin{table}[!h]
    \centering
    \begin{tabular}{lll}
    	\hline\\[-2ex]
    	& Re($\qp{1}^{2}$) & Im($\qp{1}^{2}$)\\[1ex]
    	\hline\\[-1ex]
    	Breit-Wigner & $-1.00\pm0.06$ & $\pm(0.98\pm0.03)$\\
    	Gluon DS (quenched)  & $-0.21\pm0.03$ & $\pm(0.34\pm0.02)$\\
    	Gluon lattice (quenched)  & $-0.30\pm0.07$ & $\pm(0.49\pm0.03)$\\
    	Gluon lattice (unquenched) & $-0.63 \pm 0.07$ & $\pm(0.53 \pm 0.12)$
    \end{tabular}
    \caption{\label{tab:polelocation}Location of the complex conjugate pole pair determined from the SPM reconstruction algorithm when applied on the Breit-Wigner mock data, and the gluon propagator DS~\cite{Strauss:2012dg} and lattice~\cite{Duarte:2016iko} data. The error of the input data is: $10^{-3}$ (Breit-Wigner) and $<1\%$ (gluon DS and lattice quenched, lattice unquenched).}
\end{table}

\begin{figure*}
	\includegraphics[scale=0.37]{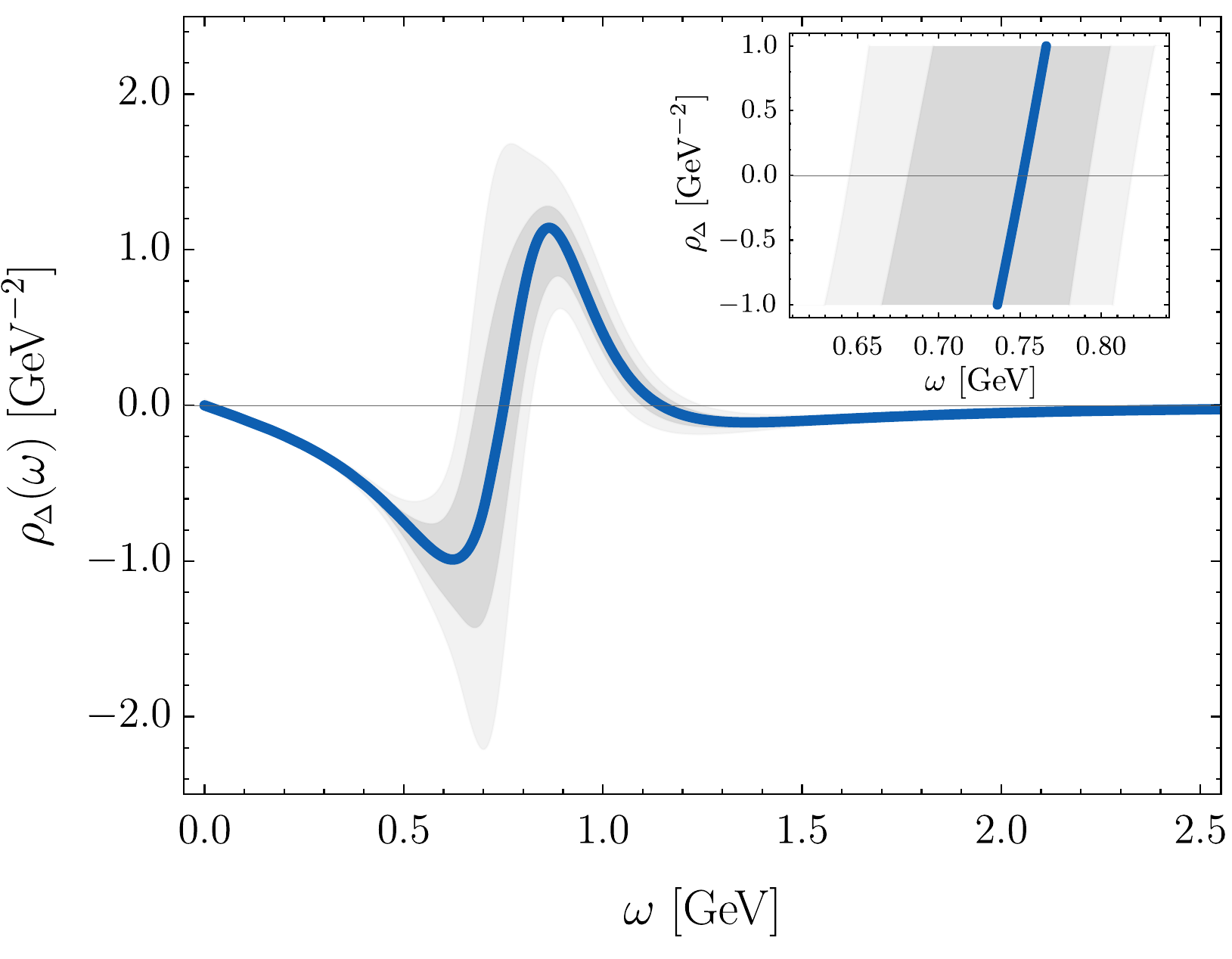}
	\includegraphics[scale=0.37]{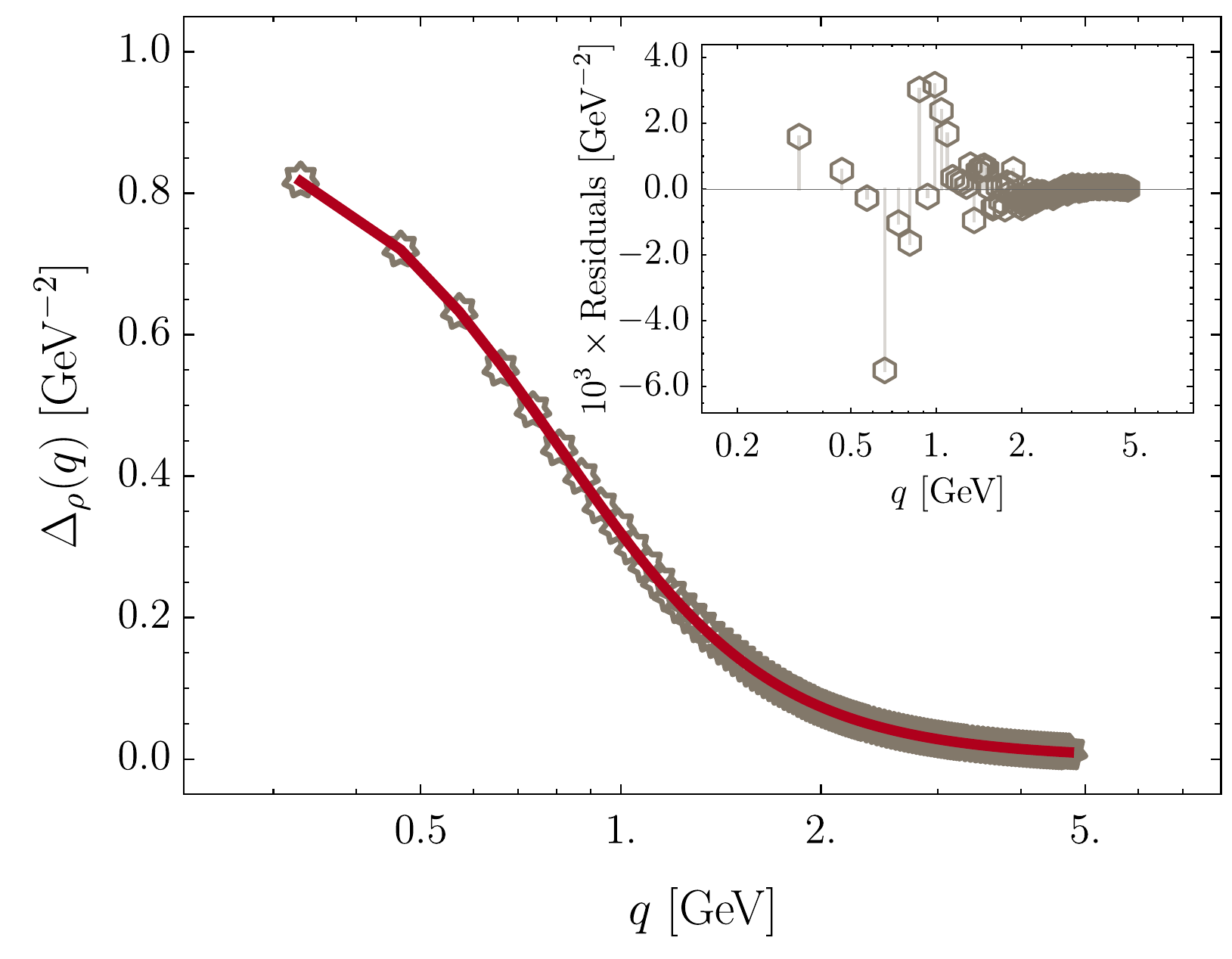}
	\includegraphics[scale=0.37]{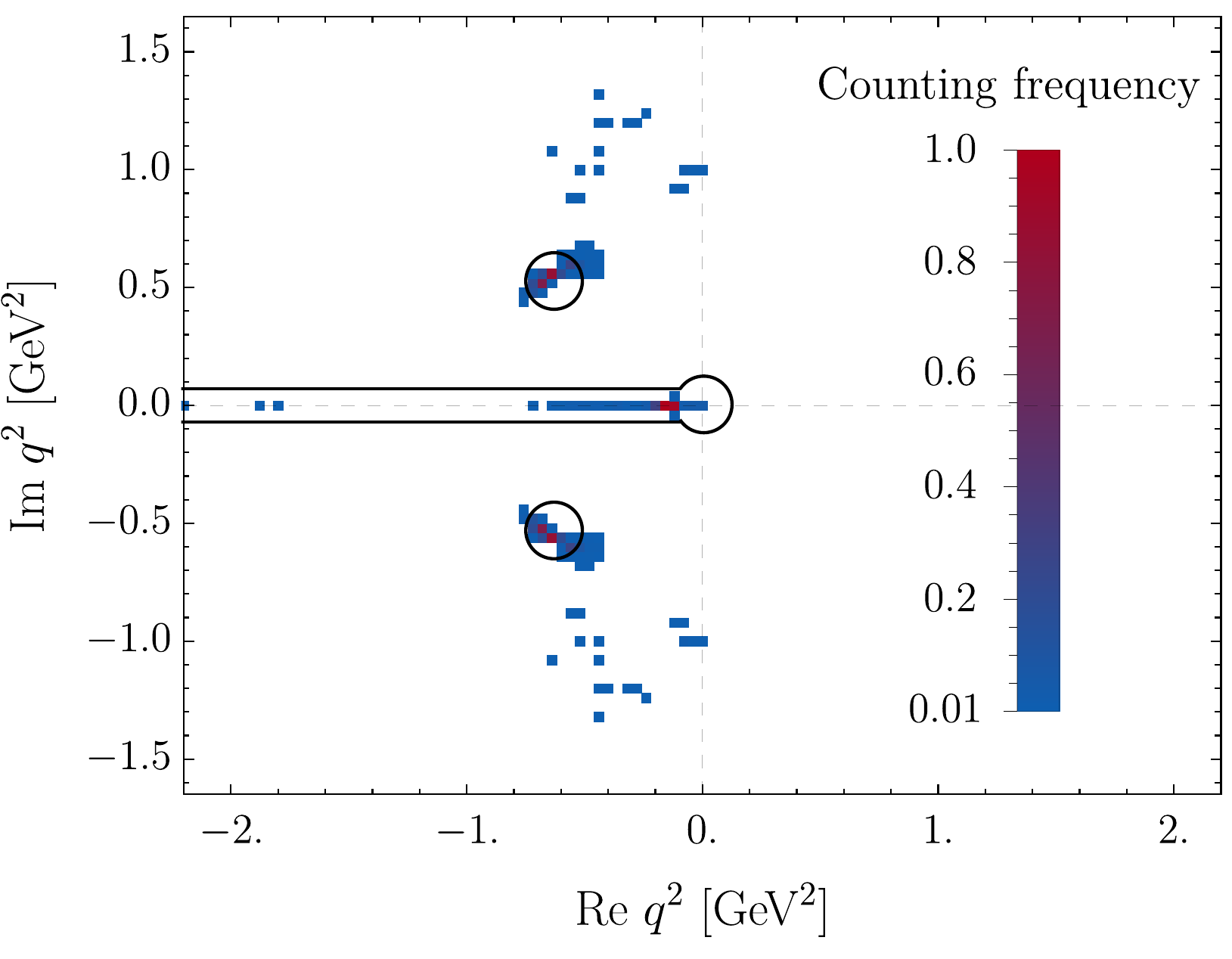}
	\includegraphics[scale=0.37]{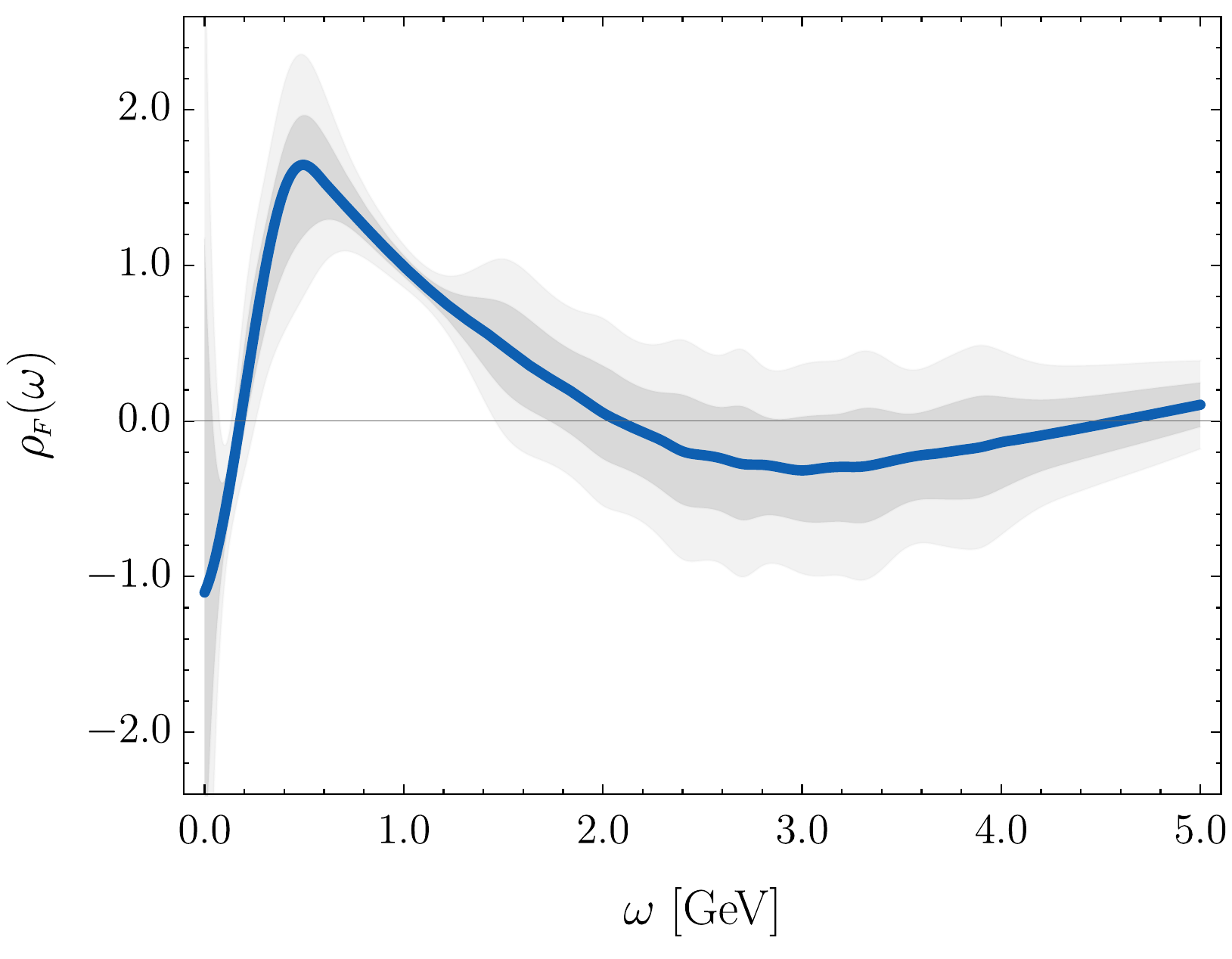}
	\includegraphics[scale=0.37]{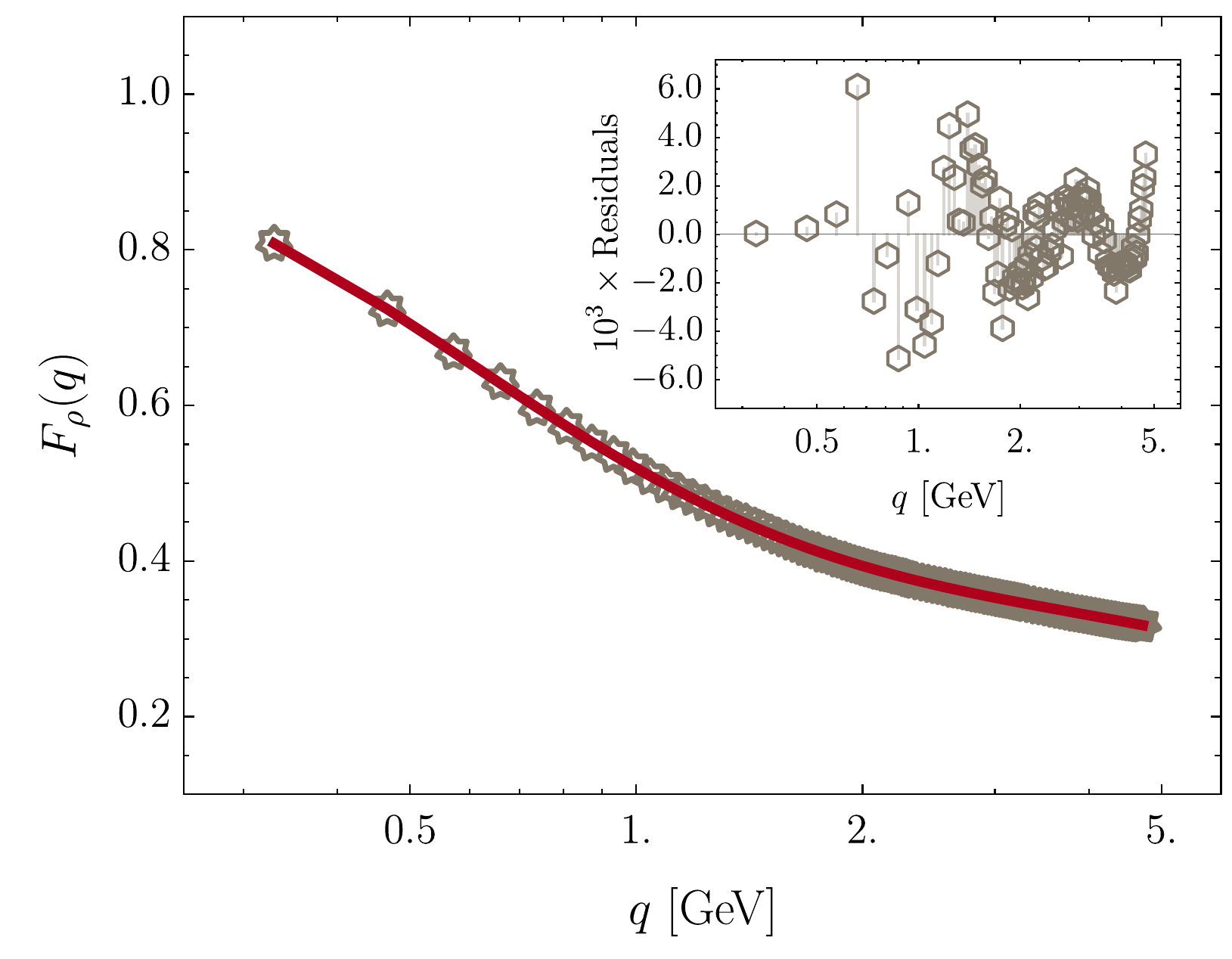}
	\includegraphics[scale=0.37]{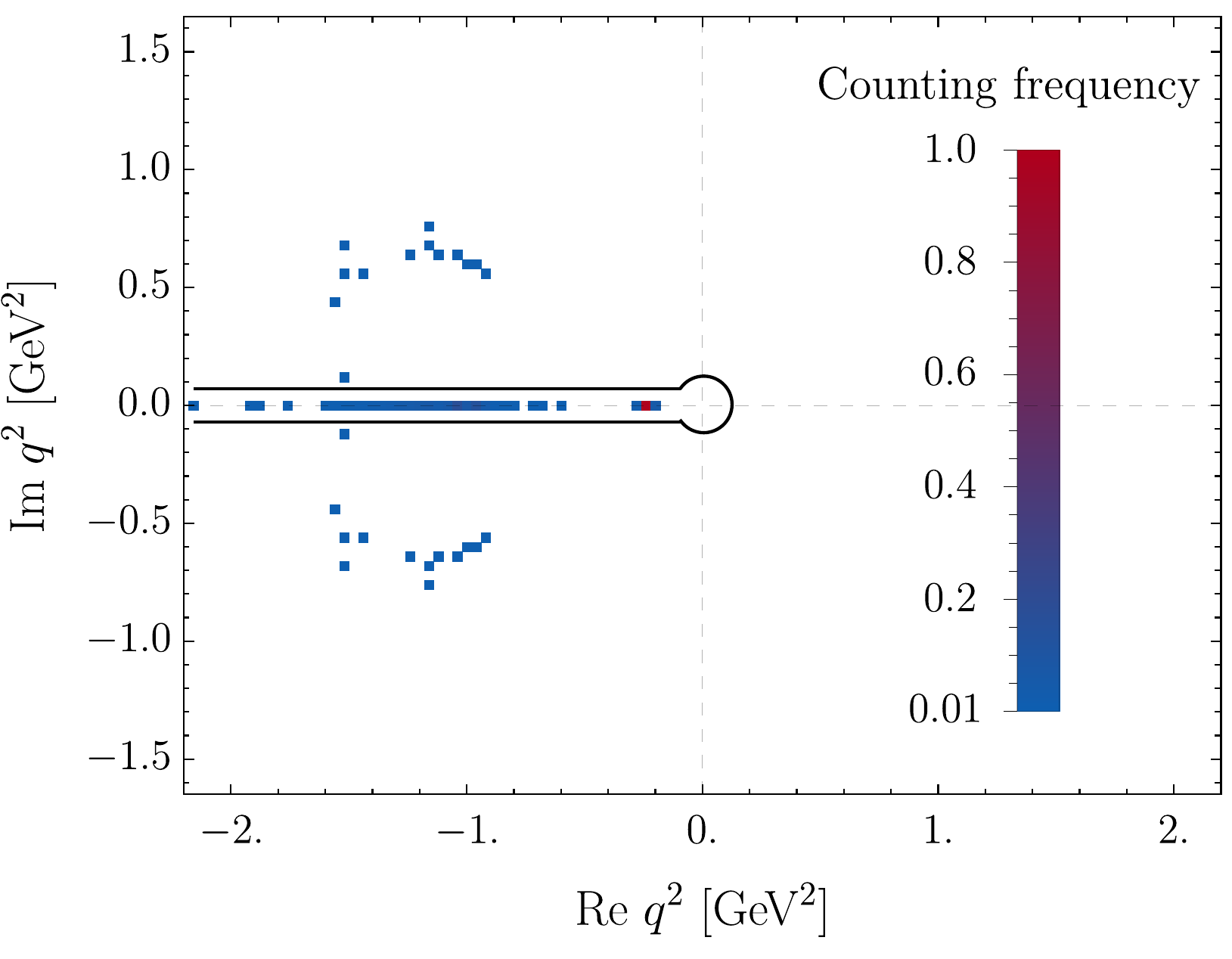}
	\caption{\label{fig:unquenched}
	As before, but for the gluon (top) and ghost (bottom) unquenched $N_f=2+1+1$ lattice data of~\cite{Ayala:2012pb}. Notice the similarities of the reconstructed spectral functions with the quenched case.}
\end{figure*}

\section{Results} 

Our results for the gluon spectral function\footnote{To ease the comparison between different results, we {\it normalize} both the DS and lattice data so that $\Delta(0)=1$ [GeV$^{-2}$] and $F(0)$=1. This should not be confused with {\it renormalization}, {\it e.g.}, it is known that it is not possible to renormalize $F$ to be unity at $q=0$ without violating symmetry identities~\cite{Aguilar:2009nf}.}~$\rgl$ are shown in~\fig{fig:gl-CF-OO}, in which the upper panels correspond to input points obtained from the numerical solution of the Dyson-Schwinger (DS) system for the gluon and ghost two-point functions~\cite{Strauss:2012dg}, whereas the lower panels use as input the $64^4$ at $\beta=6.0$ lattice results for SU(3) Yang-Mills theory of~\cite{Duarte:2016iko}. 
Qualitatively the spectral functions behave in the same way, approaching large and small frequencies from the negative region as required~\cite{Cyrol:2018xeq}, and displaying the same peak sequence; in addition, the negative contributions to the spectral function confirm the confined nature of the gluon, removing it from the theory asymptotic spectrum. The reconstructed propagators obtained from~\1eq{Lehmann} reproduce the input data well (with residuals at the $10^{-3}$ level, see insets in the middle panels). Perhaps, the most striking resemblance between the two cases can be found in the reconstructed analytic $q^2$ structure, featuring a branch cut singularity at the ghost threshold $\mathrm{Re}\, q^2\le0$ and a single pair of complex conjugate poles located at $\qp{1}^{2}\approx-0.2\pm0.3i$ GeV$^2$ (DS) and $\qp{1}^{2}\approx-0.3\pm0.5i$ GeV$^2$ (lattice), see Table~\ref{tab:polelocation} again. The presence of these poles also makes clear the quantitative difference between the two spectral functions: since the residue is larger in the lattice case with respect to the DS one, more weight for the reconstruction of the propagator is carried by the poles in the former case. Notice that these results are qualitatively different from the ones reported in~\cite{Strauss:2012dg} (DS) and~\cite{Dudal:2019gvn} (lattice), where the only analytic structure found was a branch cut. 

\begin{figure*}[!t]
	\includegraphics[scale=0.37]{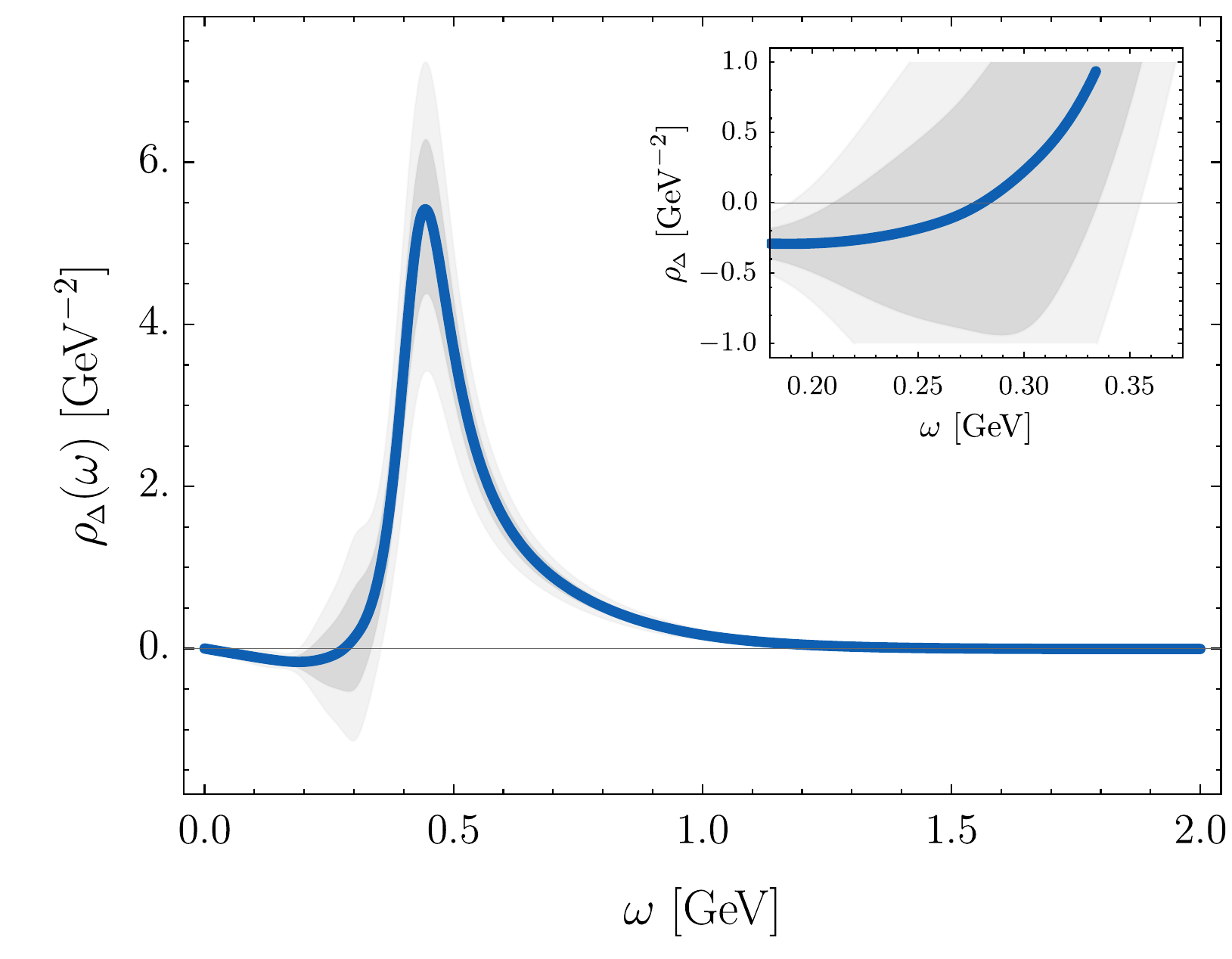}
	\includegraphics[scale=0.37]{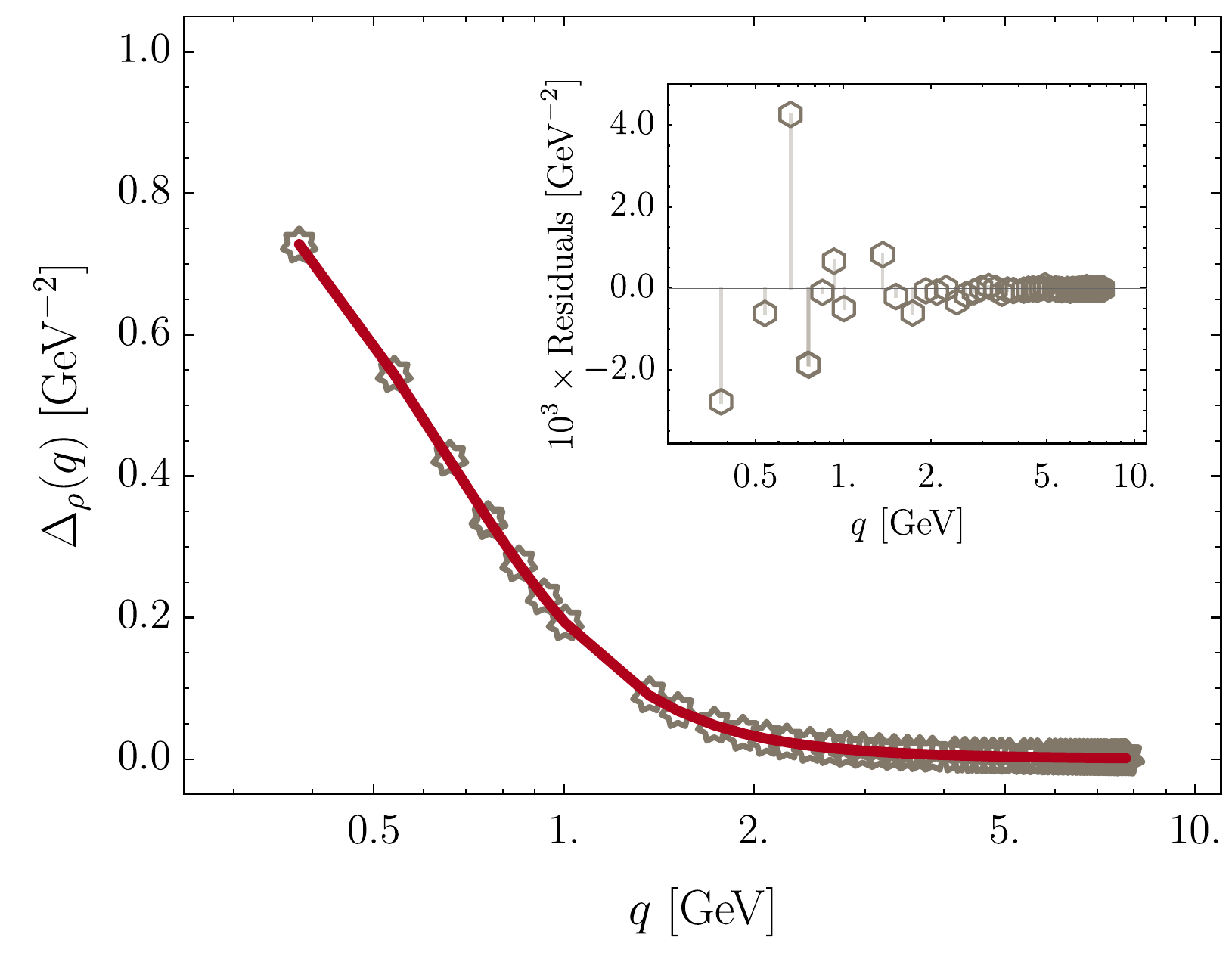}
	\includegraphics[scale=0.37]{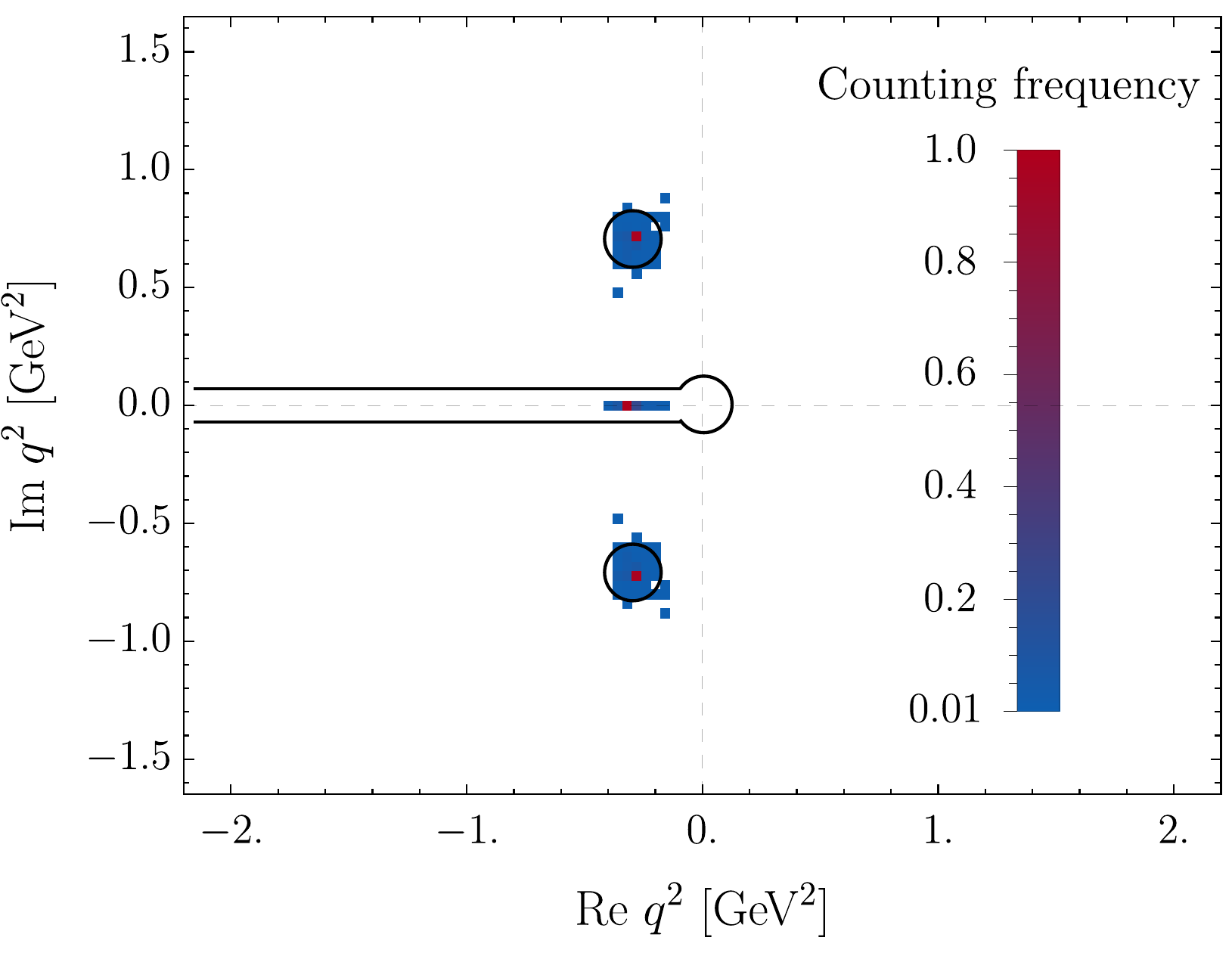}
	\caption{\label{fig:Rxi} As before but for the $R_\xi$ lattice data of~\cite{Bicudo:2015rma} at $\xi=0.5$. Poles are located at $\qp{1}^{2}\approx-0.3\pm0.7i$ GeV$^2$, whereas $\hmgl^{\,\xi=0.5}=0.28\pm0.08$.}
\end{figure*}

Knowledge of $\rgl$ allows for defining a renormalization group invariant effective gluon mass $\hmgl$. Even though this notion has proven useful, see,~{\it e.g.},~\cite{Cornwall:1981zr,Parisi:1980jy,Halzen:1992vd,Yndurain:1995uq,Szczepaniak:1995cw,Field:2001iu,Luna:2005nz,Luna:2006qp,Cornwall:2009ud,Strauss:2012dg,Binosi:2016nme}, it should be clear that $\hmgl$ is not an observable (and the gluon  not a regular massive particle); rather, $\hmgl$ coincides with the scale at which positivity violation appears. From the zero crossing of the spectral function (left panels insets if~\fig{fig:gl-CF-OO}) we can  estimate $\hmgl\approx0.4\pm0.1$ GeV (DS) and $\hmgl\approx0.6\pm0.2$ GeV (lattice). These values are compatible with available theoretical and phenomenological estimates of the RG effective gluon mass~\cite{Cornwall:1981zr,Parisi:1980jy,Halzen:1992vd,Yndurain:1995uq,Szczepaniak:1995cw,Field:2001iu,Luna:2005nz,Luna:2006qp,Cornwall:2009ud,Strauss:2012dg}, and, in particular, the estimate $\hmgl\approx0.5$ GeV obtained when constructing the process-independent QCD effective charge~\cite{Binosi:2002vk,Aguilar:2009nf} analogue of the quantum electrodynamics Gell-Mann-Low effective coupling~\cite{Binosi:2016nme}.	

We note that a qualitatively similar behavior is observed when analyzing data on the gluon propagator obtained in \cite{Cyrol:2016tym}. This data corresponds to a Landau gauge gluon propagator of the so-called scaling type, {\it i.e.},~a type of solution of the DS system that is characterized by the IR power laws $\Delta(q)\sim (q^{2})^{2\kappa-1}$ and $F(q)\sim (q^{2})^{-\kappa}$ with  $\kappa\approx0.58$ \cite{Lerche:2002ep}. While in \cite{Cyrol:2018xeq} a modified Breit-Wigner Ansatz was used to reconstruct the gluon spectral function, which by definition excludes the presence of poles in the $q^{2}$ plane, applying the SP method yields a clear signal for complex conjugate poles at $\qp{1}^{2}\approx-0.1\pm1.3i$ GeV$^2$ as well as a spectral function qualitatively similar to \fig{fig:gl-CF-OO}. 

Turning to the ghost sector, the spectral function for the dressing function~$\rghdr$ is shown in~\fig{fig:gh-CF-OO}, with the upper (lower) panels corresponding to the DS (lattice) data. We find a qualitatively similar behavior to the gluon case, albeit with larger uncertainties. Notice also that the IR asymptotics obtained from lattice data looks opposite to the one coming from the DS data; however, the latter should be considered more reliable, due to the smaller uncertainties. The analytic structure shows the presence of only a branch cut (barring some spurious poles in the lattice case due to their lower precision), starting at zero as was already the case for the gluon. This has to be expected, due to the presence of the ghost gluon vertex and the fact that the gluon, being no ordinary massive particle, carries spectral weight down to arbitrarily small energies, see~\fig{fig:gl-CF-OO}.

Similar results (shown in Fig.~\ref{fig:unquenched}) are obtained when analyzing the gluon and ghost two-point functions obtained from lattice configurations involving two light (twisted mass) degenerate quark flavours and two heavy ones~\cite{Ayala:2012pb}. The inclusion of dynamical quarks has been known to lead to a lower saturation point of the gluon propagator (see, {\it e.g.}, Fig.~1 in ~\cite{Ayala:2012pb}), which can be in turn interpreted as the gluon becoming ``more massive'' (an effect that has to be expected on theoretical grounds~\cite{Aguilar:2012rz,Aguilar:2013hoa}); on the other hand, due to the absence of a direct coupling between ghosts and quarks, the ghost is practically unaffected by the presence of the latter particles. Indeed, the gluon spectral function (see Fig.~\ref{fig:unquenched} again) reveals that the renormalization group invariant gluon mass increases to $\hmgl\approx0.75\pm0.05$ GeV; in addition, as was the case for the quenched data, one finds that the gluon spectral function is characterized by the presence of a single pair of complex conjugated poles located at $\qp{1}^{2}\approx-0.6\pm0.5i$ GeV$^2$ (see Table~\ref{tab:polelocation}), whereas the ghost  shows only the logarithmic cut.     

\section{Discussion and conclusions} 

The strength of the signal in our numerical procedure coupled to the fact that the same pole singularities appear when analyzing both the DS and lattice data, suggest that their presence is a characteristic of the gluon 2-point function rather than an artifact. This calls in turn for a thorough reassessment of DS truncation schemes as well as gauge-fixing procedures for lattice-regularized simulations (it is known in the former case that changing the vertex in the quark gap equation can significantly alter the singularity structure of the solutions, see {\it e.g.},~\cite{Burden:1991gd}). In this respect, however, we have considered the gluon propagator data of~\cite{Bicudo:2015rma}, which were obtained from lattice configurations gauge fixed in a covariant gauge at non-zero values of the gauge fixing parameter~$\xi$. The results for the case $\xi=0.5$, shown in~\fig{fig:Rxi}, are of exactly the same type of the ones found in the Landau gauge case: a pair of complex conjugate poles and a logarithmic cut\footnote{We have checked that, as expected, $\hmgl^{\,\xi=0.5}>\hmgl^{\,\xi=0}$ when the latter value is calculated within the same data set reported in~\cite{Bicudo:2015rma}.}. 

Also, in Ref.~\cite{Li:2019hyv} it has been shown that the quenched Landau gauge lattice data of~\cite{Duarte:2016iko} can be described (in the Re$(q^2)>0$ region) in terms of simple and double poles located on the real negative $q^2$-axis. These poles (which, incidentally, would not break locality) would in turn correspond to unconventional singular terms (derivatives of the $\delta$-function) present in the spectral representations of unphysical correlators, and which are permitted due to the indefinite metric characterizing covariant gauge theories (see again~\cite{Li:2019hyv} and references therein). However, the analysis of mock data generated according to these predictions, shows  that our method does correctly reconstruct such poles on the real negative $q^2$-axis; and, in particular, that it does not mistakenly reconstruct them as complex conjugate pairs, even when noise is added. Therefore, the lack of any unambiguous signal for such structures in all the cases analyzed in this work (not only the quenched Landau gauge data of~\cite{Duarte:2016iko}) strongly suggest that they are not part of a veracious spectral representation describing confined particles.

Overall, these results seem to suggest that the presence of a complex conjugate poles pair is a characteristic feature of the gluon 2-point function in covariant gauges. If confirmed, this fact would then have far-reaching consequences. At the theoretical level one would be tempted to conclude that Yang-Mills theories break locality non-perturbatively; whereas, at the phenomenological level, it would provide a new microscopic foundation for building better models of the gluon and ghost propagators leading to the development of quark-gluon kernels with the potential of describing QCD bound-states, form factors, parton distribution amplitudes and functions, at the degree of precision required by upcoming experimental efforts.
 
\section*{Acknowledgments} 
We thank C.~S.~Fischer, M.~Q.~Huber, A.~Maas, J.~M.~Pawlowski, D.~H.~Rischke, C.~D.~Roberts, L.~v.~Smekal, J.~Wambach and N.~Wink for valuable discussions; and C.~S.~Fischer and O.~Oliveira for providing respectively the DS and lattice propagator data. This work was supported by the Deutsche Forschungsgemeinschaft (DFG, German Research Foundation) – project number 315477589 – TRR 211. The work of R.-A.~T. is also supported by the BMBF under grant No.~05P18RFFCA.



\end{document}